\documentclass[11pt, oneside]{article}
\usepackage{jheppub}
\usepackage{amsmath}
\usepackage{amssymb}
\usepackage{amsfonts}
\usepackage{amsthm}
\usepackage{amsbsy}
\usepackage{array}
\usepackage{mathtools}
\usepackage[usenames,dvipsnames]{xcolor}
\usepackage{caption}
\usepackage{subcaption}
\usepackage{dsdshorthand}
\usepackage{graphicx}
\usepackage[vcentermath]{youngtab}
\usepackage{tikz}
\usepackage{booktabs,multirow}
\usepackage{listings}
\usepackage{xcolor}

\definecolor{darkblue}{rgb}{0,0,0.5}
\definecolor{darkgreen}{rgb}{0,0.3,0}
\definecolor{darkpink}{rgb}{0.4,0,0.3}
\definecolor{graygreen}{rgb}{0.3,0.5,0.3}
\definecolor{grayblue}{rgb}{0.2,0.2,0.6}
\definecolor{grayred}{rgb}{0.5,0.2,0.2}

\makeatletter
\def\@fpheader{\ }
\makeatother

\title{Bootstrapping the 3d Ising Stress Tensor}
\author{Cyuan-Han Chang$^{1,2}$, Vasiliy Dommes$^1$, Rajeev S.~Erramilli$^3$, Alexandre Homrich$^{4,5}$, Petr Kravchuk$^6$, Aike Liu$^1$, Matthew S.~Mitchell$^7$, David Poland$^7$, David Simmons-Duffin$^1$}
\affiliation{$^1$Walter Burke Institute for Theoretical Physics, Caltech, 1200 East California Blvd, Pasadena, California 91125, USA}
\affiliation{$^2$Kadanoff Center for Theoretical Physics \& Enrico Fermi Institute, University of Chicago, 933 East 56th Street, Chicago, Illinois 60637, USA}
\affiliation{$^3$Institut des Hautes \'Etudes Scientifiques, 35 Route de Chartres, 91440 Bures-sur-Yvette, France}
\affiliation{$^4$Perimeter Institute for Theoretical Physics, 31 Caroline St N, Waterloo, Ontario N2L 2Y5, Canada}
\affiliation{$^5$Kavli Institute for Theoretical Physics, University of California, 552 University Road, Santa Barbara, CA 93106, USA}
\affiliation{$^6$Department of Mathematics, King’s College London, Strand, London, WC2R 2LS, UK}
\affiliation{$^7$Department of Physics, Yale University, 217 Prospect St, New Haven, CT 06520, USA}
\emailAdd{cchang10@uchicago.edu}
\emailAdd{vdommes@caltech.edu}
\emailAdd{erramilli@ihes.fr}
\emailAdd{ahomrich@kitp.ucsb.edu}
\emailAdd{petr.kravchuk@kcl.ac.uk}
\emailAdd{aliu7@caltech.edu}
\emailAdd{matthew.mitchell@yale.edu}
\emailAdd{david.poland@yale.edu}
\emailAdd{dsd@caltech.edu}

\date{}
\abstract{We compute observables of the critical 3d Ising model to high precision by applying the numerical conformal bootstrap to mixed correlators of the leading scalar operators $\sigma$ and $\epsilon$, and the stress tensor $T_{\mu\nu}$. We obtain new precise determinations of scaling dimensions $(\Delta_\sigma, \De_\epsilon) = (0.518148806(24), 1.41262528(29))$ as well as OPE coefficients involving $\sigma$, $\epsilon$, and $T_{\mu\nu}$. We also describe several improvements made along the way to algorithms and software tools for the numerical bootstrap.}

\preprint{CALT-TH 2024-047}

\begin{document}

\maketitle
\pagenumbering{roman}
\setcounter{page}{2}
\newpage
\pagenumbering{arabic}
\setcounter{page}{1}

\setlength{\parskip}{1pt}

\section{Introduction}
\label{sec:intro}

The dream of the conformal bootstrap~\cite{Ferrara:1973yt, Polyakov:1974gs} is to use consistency conditions to solve and classify the scale-invariant structures underlying critical phenomena. A tantalizing target for this program is the critical 3d Ising model, which describes the critical point of water, uniaxial magnets, $\phi^4$ theory at strong coupling, and more. This theory has been studied for decades using the $\epsilon$-expansion~\cite{Wilson:1971dc}, high/low-temperature expansions, and Monte Carlo simulations~\cite{Pelissetto:2000ek}. The most precise results for its critical exponents have come from numerical conformal bootstrap methods~\cite{Rattazzi:2008pe, El-Showk:2012cjh, Poland:2018epd, Poland:2022qrs, Rychkov:2023wsd} applied to 4-point correlation functions of the leading $\mathbb{Z}_2$-odd and $\mathbb{Z}_2$-even scalar operators $\sigma$ and $\epsilon$~\cite{Kos:2014bka, Kos:2016ysd}. This approach has led to determinations of scaling dimensions $\Delta_{\sigma} = 1/2+\eta/2 = 0.5181489(10)$ and $\Delta_{\epsilon} = 3-1/\nu = 1.412625(10)$, along with determinations of  operator product expansion (OPE) coefficients, and clear reconstructions of leading Regge trajectories~\cite{Simmons-Duffin:2016wlq}.

However, there are several reasons why we would like to go beyond these results. Firstly, it is not clear that the allowed region of~\cite{Kos:2016ysd} can be made systematically smaller by increasing the derivative order of bootstrap functionals. Relatedly, there are certain low-lying Regge trajectories of operators that are expected theoretically \cite{Fitzpatrick:2012yx,Komargodski:2012ek,Fitzpatrick:2015qma} but not present in the extremal spectra studied in~\cite{Simmons-Duffin:2016wlq}. These include the $[\sigma T]$, $[\epsilon T]$, and $[T T]$ double-twist trajectories built out of the stress-energy tensor $T^{\mu\nu}$ (which can also be interpreted as triple-twist or quadruple-twist trajectories built out of $\sigma$). Furthermore, the spectra of~\cite{Simmons-Duffin:2016wlq} give no information about $\mathbb{Z}_2$-even operators of odd spin, parity-odd operators, or operator product expansion (OPE) coefficients involving multiple spinning operators (e.g.~those appearing in $\<TT\epsilon\>$ or $\<TTT\>$).

Meanwhile, there has been recent progress exploring the bootstrap for current~\cite{Dymarsky:2017xzb, Reehorst:2019pzi, He:2023ewx} and stress-tensor 4-point functions~\cite{Dymarsky:2017yzx}. The latter study revealed contours of a global map of the space of CFTs, parameterized by parity-even and parity-odd scalar gaps $(\Delta_+, \Delta_-)$. This map  included, among other features, a sharp discontinuity for $\De_+$ near the 3d Ising value of $\Delta_{\epsilon}$. This was a surprise, as it was not a priori clear that the stress-tensor bootstrap should know anything about the 3d Ising CFT\@. This structure was leveraged in~\cite{Dymarsky:2017yzx} to yield rough estimates of the 3d Ising $\<TTT\>$ 3-point function coefficients.

\begin{table}
\centering
\begin{tabular}{c|l|l|l|l}
& $T\s\e$ (this work) & prev.\ bootstrap & monte carlo & fuzzy sphere \\
\hline
$\Delta_\sigma$ & 0.518148806{\bf(24)} & 0.5181489(10)$^*$ \cite{Kos:2016ysd} & 0.518142(20) \cite{Hasenbusch:2021tei} & 0.524~\cite{Zhu:2022gjc}, 0.519~\cite{Fardelli:2024qla} \\
\hline
$\Delta_\epsilon$ & 1.41262528{\bf(29)} &  1.412625(10)$^*$ \cite{Kos:2016ysd} & 1.41265(13) \cite{Hasenbusch:2021tei}  &  1.414~\cite{Zhu:2022gjc}, 1.403~\cite{Fardelli:2024qla} \\
\hline
$\lambda_{\sigma\sigma\epsilon}$ & 1.05185373(11) & 1.0518537(41) \cite{Kos:2016ysd} & 1.051(1) \cite{Hasenbusch:2017yxr} & 1.0539(18) \cite{Hu:2023xak}\\
\hline
$\lambda_{\epsilon\epsilon\epsilon}$ & 1.53244304(58) &  1.532435(19) \cite{Kos:2016ysd} & 1.533(5) \cite{Hasenbusch:2017yxr} & 1.5441(23)  \cite{Hu:2023xak} \\
\hline
$c_T/c_B$ & 0.946538675(42) & 0.946534(11) \cite{El-Showk:2014dwa} & 0.952(29) \cite{Ayyar:2023wub} & 0.955(21) \cite{Hu:2023xak}  \\
\hline
$\lambda_{TT\epsilon}$ & 0.95331513(42) & 0.958(7) \cite{Poland:2023bny} &---& 0.9162(73)  \cite{Hu:2023xak} \\
\hline
$n_B$ & 0.933444559(75) & 0.933(4)$^{**}$ \cite{Dymarsky:2017yzx} & --- & --- \\
\hline
$n_F$ & 0.013094116(33) & 0.014(4)$^{**}$ \cite{Dymarsky:2017yzx} & --- & ---
\end{tabular}
\caption{\label{tab:mainresults}Bounds on scaling dimensions and OPE coefficients in the 3d Ising CFT, using the numerical bootstrap for correlators of $\sigma$, $\epsilon$, and the stress tensor $T$. Here, $c_T/c_B$ is the coefficient of the stress-tensor two-point function (the ``central charge") normalized by the free boson value, and $n_B$ and $n_F$ are OPE coefficients in the stress-tensor three-point function. The OPE coefficient $\l_{TT\e}$ is defined in section~\ref{sec:threeptstructs}. Our bounds are computed using derivative order $\Lambda=43$. Bold-face errors are rigorous. We also include comparisons to previous bootstrap calculations and the best determinations we are aware of from other numerical methods. $^*$Note: the errors in~\cite{Kos:2016ysd} have been slightly underestimated, see footnote~\ref{foot:newse}. $^{**}$Note: the results for $(n_B,n_F)$ in \cite{Dymarsky:2017yzx} assume a parity-odd scalar gap $\De_-\geq 10$. The weaker assumption $\De_-\geq3$ gave $(n_B, n_F) = (0.919(18), 0.027(18))$.}
\end{table}

\begin{table}
    \centering
    \begin{tabular}{c|l|l}
         & $\eta$ & $\nu$ \\
        \hline
        $T\s\e$ (this work)  & 0.036297612{\bf(48)} &0.62997097{\bf(12)}\\
        \hline
        $\s\e$ bootstrap  \cite{Kos:2016ysd} & 0.0362978(20) & 0.6299710(40)\\
        \hline
        monte carlo \cite{Hasenbusch:2021tei}  & 0.036284(40) & 0.62998(5) 
    \end{tabular}
    \caption{\label{tab:exponentcomparison}Bounds on the critical exponents, derived from the scaling dimensions in Table \ref{tab:mainresults} for the reader's convenience. Bold-face errors are rigorous.}
\end{table}
\begin{figure}
\hspace{-0.5cm}
\begin{tikzpicture}
\node at (0,0) {\includegraphics[width=8cm]{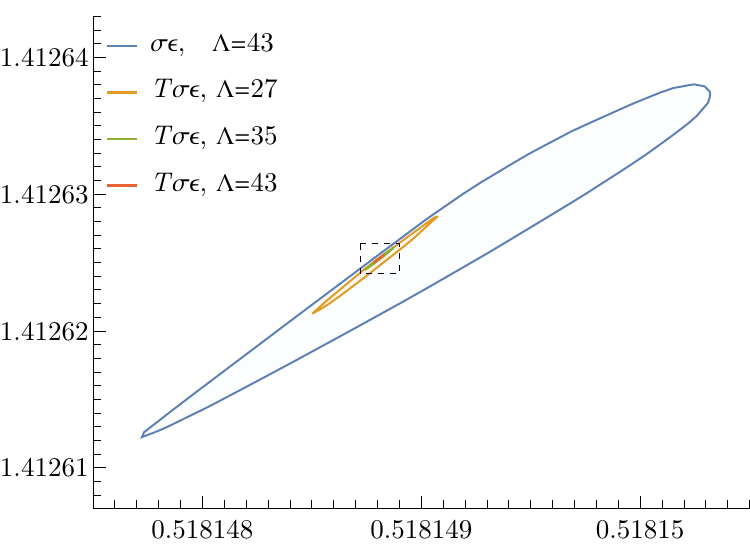}};
\node at (7.9,-0.05) {\includegraphics[width=8cm]{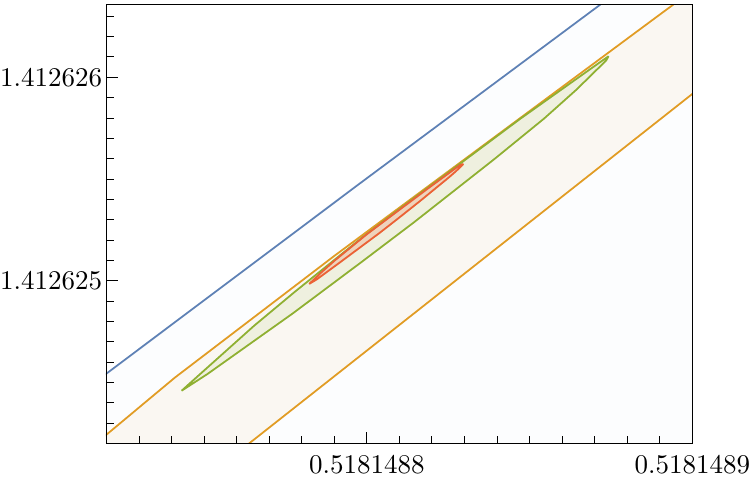}};
\draw[dotted] (-0.165,-0.012) -- (5.04,-2.215);
\draw[dotted] (0.255,-0.012) -- (5.04,-0.987381); 
\draw[dotted] (-0.165,0.305) -- (5.04,2.475);
\draw[dotted] (0.255,0.305) -- (5.04,1.2828); 
\node at (4.4,-2.5) {$\De_\s$};
\node at (-2.9,3) {$\De_\e$};
\end{tikzpicture}
\caption{\label{fig:islands} Allowed regions in $(\De_\s,\De_\e)$-space determined from the $T\s\e$ system of correlators at $\Lambda=27,35,43$ (orange, green, and red), together with the allowed region determined from the $\s\e$ system at $\Lambda=43$ (blue) \cite{Kos:2016ysd}.\protect\footnotemark\ On the left, we show a wider view where all the regions are visible. On the right, we zoom in on the $T\s\e$ regions at $\Lambda=35$ and $\Lambda=43$. At $\Lambda=43$, the $T\s\e$ allowed region is roughly $40$-$50\x$ smaller than the $\s\e$ allowed region.}
\end{figure}

A natural continuation of these works is to consider a mixed-correlator system involving all nonvanishing 4-point functions of $\sigma$, $\epsilon$, and $T^{\mu\nu}$. While such a large system comes with technical challenges, they are not insurmountable and we will see that combining these constraints allows us to make precise predictions for CFT data involving the stress tensor, opening up a new era of precision in the numerical study of the 3d Ising CFT\@. Along the way we have significantly improved several software tools and algorithms used for the numerical bootstrap. For the impatient reader, a concise summary of our new results for scaling dimensions and OPE coefficients is provided in table~\ref{tab:mainresults}. The corresponding bounds on the critical exponents $\eta, \nu$ are shown in table~\ref{tab:exponentcomparison}. We also provide a plot of the allowed island in $(\De_\s,\De_\e)$-space (which is about 50 times smaller than the previous state of the art) in figure~\ref{fig:islands}.

Experience has shown that bootstrap bounds tend to converge as one increases the derivative order, but these converged bounds are {\it not\/} exactly saturated by physical theories (though they may be nearly saturated). In other words, there is a limit to the strength of bounds extracted from a fixed, finite set of crossing equations. In the context of the hyperbolic bootstrap \cite{Bonifacio:2021msa,Bonifacio:2021aqf,Kravchuk:2021akc}, there are essentially rigorous arguments that this non-saturation occurs \cite{Radcliffe:2024jcg}. Thus, if the bootstrap is to become a practical tool for solving a theory, we {\it must\/} explore larger and larger systems of crossing equations. We view our results as more evidence (along with \cite{Kos:2014bka,Chester:2019ifh,Chester:2020iyt,Erramilli:2022kgp}) that this is a fruitful thing to do! 

This paper is organized as follows. In section~\ref{sec:setup} we describe the basic setup of the bootstrap problem, including the analysis of the 3-point structures, Ward identities and conservation constraints, and crossing equations. In section~\ref{sec:implementation} we describe our numerical implementation of the bootstrap problem, along with various improvements to software tools that we have made. In section~\ref{sec:certifying} we describe a new rigorous approach to certifying disallowed points in the context of scanning over OPE coefficients. Section~\ref{sec:results} describes the results of our computations, including new bounds from the $\<TTTT\>$ single-correlator system and precise computations of 3d Ising data using the mixed correlator $\{\sigma, \epsilon, T\}$ system. We conclude in section~\ref{sec:conclusions} and give details of our setup and implementation in the appendices.

\section{Setup}
\label{sec:setup}

The non-trivial four-point functions in the $\{\sigma, \epsilon, T\}$ system are 
\be
\{\<\s\s\s\s\>,\<\s\s\e\e\>,\<\e\e\e\e\>,\<\s\s\e T\>,\<\e\e\e T\>,\<\s\s TT\>,\<\e\e TT\>, \<\e TTT\>,\<TTTT\>\}.
\ee
They are constrained by conformal, parity, and permutation symmetries. Moreover, the correlators involving the stress tensor must be conserved and satisfy Ward identities. In this section, we describe how these constraints are imposed and determine a linearly-independent set of crossing equations. We also study the three-point functions relevant for constructing the conformal block expansions of the above four-point functions. We generally follow the formalism developed and described in~\cite{Kravchuk:2016qvl, Dymarsky:2017yzx, Dymarsky:2013wla, Erramilli:2019njx, Erramilli:2020rlr, Erramilli:2022kgp}, with a few additional considerations necessary for numerical stability.

\footnotetext{\label{foot:newse}The $\s\e$ allowed region shown in this figure was re-computed for this work and is a more accurate version of the allowed region from~\cite{Kos:2016ysd}. Compared to \cite{Kos:2016ysd}, we have benefitted from the OPE cutting surface algorithm \cite{Chester:2019ifh}, improvements in computational efficiency, and more powerful computational resources. Our $\s\e$ allowed region is slightly larger than the one in~\cite{Kos:2016ysd} but this does not significantly affect any of their results. We thank Slava Rychkov and Ning Su for discussions that motivated us to perform this calculation.}

\subsection{Three-point structures}
\label{sec:threeptstructs}

\begin{table}[t]
\centering
\begin{tabular} {c | c | c | c }
\hline
OPE & spin & $\Z_2$  & \#structures$^\textrm{parity}$ \\
\hline
$\s \x \s$ & even & even & $1^+$ \\
\hline
$\e \x \e$ & even & even & $1^+$ \\
\hline
$\s \x \e$ & $j\geq 0$ & odd & $1^+$ \\
\hline
\multirow{2}{*}{$\s\x T$} & \multicolumn{2}{|c|}{$\sigma$} & $1^{+}$ \\
\cline{2-3}
 & $j\geq 2$ & odd & $1^{+} + 1^{-}$ \\
\hline
\multirow{2}{*}{$\e\x T$} & \multicolumn{2}{|c|}{$\e$} & $1^{+}$ \\
\cline{2-3}
 & $j\geq 2$ & even & $1^{+} + 1^{-}$ \\
\hline
\multirow{5}{*}{$T\x T$} & $j=0$ & even & $1^{+} + 1^{-}$ \\
\cline{2-3}
 & $j=2$ & even & $1^{+} + 1^{-}$ \\
\cline{2-3}
 & \multicolumn{2}{|c|}{$T$} & $2^{+}$ \\
 \cline{2-3}
 & $j\geq 4$ even & even & $2^{+} + 1^{-}$ \\
 \cline{2-3}
 & $j\geq 5$ odd & even & $1^{-}$ \\
 \hline
\end{tabular}
\caption{The allowed operators and number of tensor structures in each OPE between $\s,\e$, and $T$. The number of structures is labeled by parity: $n^+$ stands for $n$ parity-even structures, while $n^-$ stands for $n$ parity-odd structures.}
\label{tab:OPE}
\end{table}

We have studied all the OPEs between $\s,\e,T$ and computed their $q$-basis~\cite{Kravchuk:2016qvl} three-point structures satisfying conformal invariance, invariance under space parity and permutations, and conservation of $T$. With the exception of the latter, these constraints can be straightforwardly implemented using the formalism described in~\cite{Kravchuk:2016qvl}. The conservation of $T$ can be imposed by converting to embedding space bases~\cite{Costa:2011mg,Costa:2011dw}; we, however, chose to impose it directly in the $q$-basis, which can be done following the discussion in section 4.3 of~\cite{Cuomo:2017wme}. Since these manipulations are standard, we do not describe them in detail and refer the interested reader to the above references.

The results for the representations of primary operators $\cO$ operators appearing in various OPEs and the respective numbers of tensor structures are summarized in table~\ref{tab:OPE}. Note that the counting of conserved structures generally splits into two cases, generic $\cO$ and special $\cO$. This is why the $\s$ operator in the $\s\x T$ OPE, $\e$ in the $\e \x T$ OPE, and $T$ in the $T\x T$ OPE are treated separately. For example, conservation and Ward identities imply that $\s$ is the only scalar that can appear in the $\s\x T$ OPE.

The last step is to convert the tensor structures from the $q$-basis to the $\SO(3)_r$ basis used by {\tt blocks\_3d}~\cite{Erramilli:2019njx,Erramilli:2020rlr}, a program that can compute arbitrary four-point conformal blocks in 3d CFTs. This can be done by writing each $q$-basis structure as a linear combination of the $\SO(3)_r$ basis structures with the coefficients given by products of the Clebsch-Gordan coefficients; we refer the reader to \cite{Erramilli:2019njx} for more details. The explicit expressions of all the three-point structures in the $\SO(3)_r$ basis are given in the ancillary file  \texttt{three\_point\_structures.m}. We have also checked that our structures agree with previous literature.\footnote{In particular, the $\<TT\cO\>$ structures are linear combinations of the differential basis expressions used in the code of \cite{Dymarsky:2017yzx}. The actual expressions listed in appendix A of \cite{Dymarsky:2017yzx} for the $\langle TTT \rangle$ and parity-odd $\langle TTO\rangle$ structures contain typos. See the ancillary file  \texttt{three\_point\_structures.m} for the corrected explicit expressions.} The full list of three-point structures corresponding to the possible OPE coefficients in the $T\s\e$ system are listed in appendix~\ref{app:threept}.

Our definition of the OPE coefficient $\l_{TT\e}$ is slightly modified in order to match the conventions of \cite{Poland:2023vpn}, namely:
\be
\label{eq:tte}
&\<\hat T(X_1,Z_1) \hat T(X_2,Z_2) \e(X_3)\> \nn\\
&= \l_{TT\e}\frac{1}{X_{12}^{\frac{6-\De_\e}{2}+2} X_{23}^{\frac{\De_\e}{2}} X_{31}^{\frac{\De_\e}{2}}}\p{V_1^2 V_2^2 + \frac{4 (\Delta_\epsilon -4)}{\Delta_\epsilon +2} H_{12}V_1 V_2 + \frac{2(\Delta_\e ^2-6 \Delta_\e +6)}{\Delta_\e  (\Delta_\e +2)} H_{12}^2} \nn\\
&= -\l_{TT\e}\frac{16\De_\e}{2 + \De_\e}\ \texttt{ScalarParityEven}_{\De_\e},
\ee
where $V_i$ and $H_{ij}$ are standard embedding space structures \cite{Costa:2011mg}, and $\hat T$ is the stress-tensor with {\it unit-normalized\/} two-point function, defined in section~\ref{sec:ward} below. In the $\SO(3)_r$ basis, the structure $\texttt{ScalarParityEven}_\De$ is defined as
\be
\texttt{ScalarParityEven}_\De &= -\frac{(\Delta -5) (\Delta -3)}{4 \sqrt{5} \Delta ^2} |0,0\>+\frac{(5-\Delta )}{2 \sqrt{14} \Delta } |2,2\>-\frac{(\Delta +2)}{4 \sqrt{70} \Delta } |4,4\>,
\ee
where $|j_{12},j_{123}\>$ are $\SO(3)_r$ basis elements.
($\texttt{ScalarParityEven}_\De$ is also given in a machine-readable format in the ancillary file.) The expressions for (\ref{eq:tte}) under permutations of the 3 operators are given in (\ref{eq:externalopes}).

\paragraph{Linear independence of structures} We emphasize that it is important to ensure that three-point structures for a given channel are linearly-independent for {\it all\/} allowed values of $\De$.\footnote{We observed empirically that one only needs to ensure linear independence for the values of $\De$ allowed by unitarity and the gap assumptions. Presumably this happens because in {\tt SDPB} the conformal blocks are eventually sampled at a large number of allowed values of $\De$ and polynomially interpolated. This interpolation will very accurately reproduce any degeneracies in the allowed region of $\De$, but will likely break any degeneracies outside of this region where the polynomial interpolation is of poor quality.}
 Without care, it can happen that the space spanned by a na\"ive choice of conserved structures is degenerate at specific values of \(\De\). This occurs, for example, if one solves the conservation equations in {\tt Mathematica} in a generic basis.

These comments include the special case where structures have different asymptotic behaviors in the limit of \(\Delta\to\infty\). When this happens, the subleading structures effectively become zero, so we have a linear dependence at $\Delta=\infty$. To address this issue, we choose to normalize our structures so that they all grow as $O(1)$ as $\De \to \oo$.

When we first set up the $\<TTTT\>$ system, we hadn't appreciated the significance of this requirement. We found to our frustration that {\tt SDPB} never terminated with {\tt dual feasible jump detected}. We eventually realized that any finite linear combination of functionals in our chosen basis would always lead to a degenerate matrix at some $\De$ (including $\De=\oo$), so the solver could never find a solution where all matrices are strictly positive definite, a prerequisite for a dual feasible jump. Dual jumps are essential for efficient checkpointing \cite{Go:2019lke}, and consequently for efficient use of the OPE cutting surface algorithm \cite{Chester:2019ifh}. Thus, ensuring linear independence of three-point structures is a requirement for this kind of numerical study.

\subsection{Ward identities}
\label{sec:ward}

Some three-point functions involving the stress tensor satisfy Ward identities, which can be seen by integrating $T$ over a surface to obtain conformal generators. For example, the dilatation operator on a surface $\Sigma$ is given by
\be
D(\Sigma) = -\int_{\Sigma}dS_{\mu}x_{\nu}T^{\mu\nu}.
\ee
For a three-point function $\<\cO\cO T\>$, choosing $\Sigma$ to surround one of the $\cO$'s then gives a Ward identity of the form
\be\label{eq:Ward_identity}
-\int_{\Sigma}dS_{\mu}x_{\nu}\<\cO\cO T^{\mu\nu} \> = \De_{\cO} \<\cO\cO\>,
\ee
which relates the OPE coefficients $\l_{\cO\cO T}$ to $\De_{\cO}$.

The two-point function of stress tensors can be parametrized as
\be
\<TT\> &= c_T  \frac{H_{12}^2}{X_{12}^5},
\ee
where $H_{12}$ and $X_{12}$ are standard embedding-space structures \cite{Costa:2011mg}, and $c_T$ is called the ``central charge".\footnote{This is standard but imprecise terminology, since unlike in 2d, $c_T$ is not actually the central charge of a symmetry algebra.} The three-point function of stress tensors takes the form
\be
\<TTT\> &= n_B \<TTT\>_B + n_F\<TTT\>_F,
\ee
where $\<TTT\>_B$ and $\<TTT\>_F$ denote three-point functions in the theories of a free real scalar and a free Majorana fermion, respectively. The Ward identity (\ref{eq:Ward_identity}) then implies that $c_T = c_B n_B + c_F n_F = c_B(n_B+n_F)$, where
\be
c_B = c_F = \frac{3}{2(4\pi)^2}
\ee
are the central charges of the free scalar and free Majorana fermion theories. The average null energy condition constrains $n_B$ and $n_F$ to be nonnegative: $n_B\geq 0$ and $n_F\geq 0$, with saturation if and only if the theory is free \cite{Hofman:2008ar,Hofman:2016awc,Hartman:2016lgu,Faulkner:2016mzt}.

In our numerical setup, it is useful to work in conventions where the two-point function of the stress tensor is unit-normalized. Let us therefore define the normalized stress tensor
\be
\hat T \equiv \frac{T}{\sqrt{c_T}},\qquad \<\hat T \hat T\> = \frac{H_{12}^2}{X_{12}^5}.
\ee
Three-point functions of $\hat T$ are given by
\be
\<\hat T \hat T \hat T\> &= \l_B \<\hat T\hat T\hat T\>_B + \l_F\<\hat T\hat T\hat T\>_F,
\ee
where $\<\hat T \hat T \hat T\>_{B,F}$ are the corresponding 3-point functions in free theories, and
\be
\l_B \equiv \frac{n_B}{(c_T/c_B)^{3/2}}=\frac{n_B}{(n_B+n_F)^{3/2}},\qquad \l_F \equiv \frac{n_F}{(c_T/c_F)^{3/2}}=\frac{n_F}{(n_B+n_F)^{3/2}}.
\ee
Note that these normalized coefficients satisfy
\be
\label{eq:normalizedcoeffidentity}
\l_B + \l_F &= \frac{1}{(n_B+n_F)^{1/2}} = \frac{1}{\sqrt{c_T/c_B}}.
\ee

Three-point correlators involving two scalars and a stress tensor are also constrained by Ward identities. For any scalar $\f$, we have
\be
\label{eq:nicewardforscalars}
\<\f\f \hat T\> &= -\frac{\De_\f}{\sqrt{c_T/c_B}} |0,2\> = -\De_\f(\l_F + \l_B) |0,2\>,
\ee
where $|0,2\>$ denotes an $\SO(3)_r$ structure \cite{Erramilli:2019njx,Erramilli:2020rlr}. On the right-hand side, we used the identity (\ref{eq:normalizedcoeffidentity}). The fact that $\l_B,\l_F$ appear linearly in Ward identities is convenient for our implementation: we can treat them like any other independent OPE coefficients, with the subtlety that they appear in multiple three-point correlators.

For the rest of the paper we implicitly assume unit-normalized stress tensors and therefore dispense with the hat notation.

\subsection{Four-point structures and crossing equations}
\label{subsec:4pt}

Next, we must establish a complete set of independent crossing equations for the four-point correlators in the  $T \sigma \epsilon $ system.

\paragraph{Four-point tensor structures} To describe four-point tensor structures we use the $q$-basis of~\cite{Kravchuk:2016qvl}, as it does not suffer from redundancies which plague embedding-space bases~\cite{Costa:2011mg, Dymarsky:2013wla}. A $q$-basis four-point tensor structure is denoted by $[q_1q_2q_3q_4]^\pm$, where $q_i\in \{-\ell_i,-\ell_i+1,\dots,+\ell_i\}$ and $\ell_i$ is the spin of the operator at position $i$. The sign $\pm$ indicates the charge of the tensor structure under exchange of the cross-ratios $z\leftrightarrow\bar z$. 

A general four-point function can be expanded as
\be
	\<\cO_1\cO_2\cO_3\cO_4\>=\sum_{\{q_i\},\pm} [q_1q_2q_3q_4]^\pm g^{[q_1q_2q_3q_4]^\pm}_{\cO_1\cO_2\cO_3\cO_4}(z,\bar z).
\ee
The functions $g^{[q_1q_2q_3q_4]^\pm}(z,\bar z)$ satisfy $g^{[q_1q_2q_3q_4]^\pm}_{\cO_1\cO_2\cO_3\cO_4}(\bar z,z)=\pm g^{[q_1q_2q_3q_4]^\pm}_{\cO_1\cO_2\cO_3\cO_4}(z,\bar z)$. We refer the reader to~\cite{Kravchuk:2016qvl} for details. We also define coordinates $x,t$ by\footnote{Note that these are different from the $x,t$-coordinates defined by {\tt blocks\_3d}~\cite{Erramilli:2020rlr}.}
\be
	z=x+it,\quad \bar z=x-it.
\ee

Permutation symmetries of the four-point functions split into crossing transformations and kinematic permutations. The latter are permutations that preserve the cross-ratios, see~\cite{Kravchuk:2016qvl} for details. The group $\Pi$ of kinematic permutations is given by
\begin{align}
\Pi_{\langle TTTT\rangle} &= \{\text{id}, (12)(34),(13)(24),(14)(23)\},\\
\Pi_{\langle TT\sigma\sigma\rangle} &= \Pi_{\langle TT\epsilon\epsilon\rangle} = \{\text{id}, (12)(34)\},\\
\Pi_{\langle T\epsilon\sigma\sigma\rangle} &= \Pi_{\langle TTT\epsilon\rangle} = \Pi_{\langle T\epsilon\epsilon\epsilon\rangle} = \{\text{id}\}.
\end{align}
with analogous expressions for other orderings. Any $q$-basis structure can be symmetrized under $\Pi$, and we denote symmetrization of $[q_1q_2q_3q_4]^\pm$ by $\<q_1q_2q_3q_4\>^\pm$. Four-point functions can now be written as 
\be\label{eq:genearlfour-pt}
\<\cO_1\cO_2\cO_3\cO_4\>=\sum_{\{q_i\}/\Pi,\pm} \<q_1q_2q_3q_4\>^\pm g^{\<q_1q_2q_3q_4\>^\pm}_{\cO_1\cO_2\cO_3\cO_4}(z,\bar z).
\ee
Finally, we can account for space parity by restricting the sum to those $q$ for which the structure $\<q_1q_2q_3q_4\>^\pm$ has the desired parity. The parity is always even in our setup, which restricts the sum $\sum_i q_i$ to be even.

We do not provide the explicit lists of the more than one hundred four-point tensor structures. Instead, we provide some counting data in table~\ref{tab:counting_4pt_structures}.

\paragraph{Conservation} The four-point functions must satisfy the conservation equations for $T$, 
\be
	\<\ptl_\mu T^{\mu\nu}\cdots\>=0.
\ee
These are differential equations for each four-point function involving $T$, which have many components due to the index \(\nu\)  and the spin indices of other insertions. These equations heavily constrain the functional degrees of freedom of our four-point functions. As we explain below, it is necessary to properly account for these functional degrees when choosing which crossing equations to include in a bootstrap computation. Due to the conformal invariance of the conservation operator \(\ptl_\mu T^{\mu\nu}\), the left-hand side can be expanded in a basis of conformally-invariant four-point tensor structures for a correlator in which $T$ is replaced by a vector operator.

Using the expansion~\eqref{eq:genearlfour-pt} the conservation equations can then be converted to systems of first-order partial differential equations for the functions $g^{\<q_1q_2q_3q_4\>^\pm}_{\cO_1\cO_2\cO_3\cO_4}$.\footnote{Since the $q$-basis is defined using a choice of conformal frame, one again needs to use conformal invariance as described in section 4.3 of~\cite{Cuomo:2017wme}.}  Denoting the full set of such coefficient functions as a vector $\vec{g}_{\cO_1 \cO_2 \cO_3 \cO_4 }$, one can express the collection of differential equations as
\be
\left(M_1 \ptl_t + M_2 \ptl_x + M_3\right)\cdot \vec{g}_{ \cO_1 \cO_2 \cO_3 \cO_4 } = 0, \label{eq:finalcons}
\ee
where $M_i$ are $x,t$-dependent matrices. If $M_1$ had a left inverse, it would be possible to solve this equation given some initial data on the $t=0$ slice. Instead, $M_i$ has a non-trivial kernel, so in general we must specify some number of components of $\vec g_{\cO_1\cO_2\cO_3\cO_3}$ in the bulk (i.e. over all $x,t$) in order to solve for the remaining components in terms of $t=0$ initial data. We refer to the former as ``bulk'' degrees of freedom---these are two-variable degrees of freedom in the four point function which are not constrained by the conservation equations, and which will therefore enter our bootstrap setup.

The initial data at $t=0$ must satisfy further constraints. Let $N$ be a matrix whose rows span the left null space of $M_1$. Multiplying~\eqref{eq:finalcons} by $N$ from the left, we find 
\be
\left(NM_2 \ptl_x + NM_3\right)\cdot \vec{g}_{ \cO_1 \cO_2 \cO_3 \cO_4 } = 0,
\ee
which at $t=0$ can be seen as a constraint that analogously reduces some of the initial data from ``line'' degrees of freedom to ``point'' degrees of freedom.

Yet further but nonetheless important constraints arise at $t=0$ as this is a locus of enhanced symmetry: when $t=0$ the four operators lie all on one circle. It can be shown that this forces some low-order $t$-derivatives of $\vec{g}_{ \cO_1 \cO_2 \cO_3 \cO_4 }$ to vanish at $t=0$: see appendix A of~\cite{Kravchuk:2016qvl}. 

The complete enumeration of bulk, line, and point degrees of freedom is precisely the set of functional degrees of freedom on which we impose our crossing equations. In practice, the bulk degrees of freedom enter more or less as normal with the usual list of derivatives in our expansion, while the line and point degrees of freedom enter with just the derivatives along \(z=\bar{z}\) and without any derivatives, respectively. A summary of the various degrees of freedom in each of the four-point functions can be found in table~\ref{tab:counting_4pt_structures}. The complete summary of crossing equations is provided in appendix~\ref{sec:appendix_crossing}.

The above discussion of conservation has been quite schematic as our goal was merely to highlight the key issues. For details, we refer the reader to~\cite{Dymarsky:2017yzx}; the analysis here is a straightforward if somewhat tedious generalization.

\begin{table}
\centering
\begin{tabular}{l||c|c|c||c|c|c}
 &conformal & parity & permutation &bulk & line & point \\
\hline
{$\langle TTTT \rangle $} & 625 & 313 & 97 & 5 & 9 & 8\\
{$\langle TTT\epsilon \rangle $} & 125 & 63 & 63 & 4 & 3 & 3 \\
{$\langle TT\epsilon \epsilon \rangle $} & 25 & 13 & 9 & 2 & 1 & 0 \\
{$\langle TT\sigma \sigma \rangle $} & 25 & 13 & 9 & 2 & 1 & 0 \\
{$\langle T \epsilon \epsilon \epsilon \rangle $} & 5 & 3 & 3 & 1 & 0 & 0\\
{$\langle T \epsilon \sigma \sigma \rangle $} & 5 & 3 & 3 & 1 & 0 & 0
 \end{tabular}
\caption{\label{tab:counting_4pt_structures} The number of tensor structures contributing to spinning correlators of the  $T \sigma\epsilon$ system as conformal symmetry, parity invariance,  permutation invariance, and stress-tensor conservation are imposed. For the latter, we count bulk, line, and point degrees of freedom separately.}
\end{table}

\paragraph{Linear independence of crossing equations}

We must emphasize that it is strictly necessary to reduce the set of crossing equations to the minimal set of independent equations. Na\"ively, one might expect that imposing additional crossing equations might at most slow down numerical computations, but ultimately would lead to equivalent and correct bounds on CFT data. In practice, we found that the presence of spurious crossing equations lead to nearly-flat directions in the SDP, begetting numerical instabilities which ultimately can lead to erroneously ruling out part of theory space. In fact, with typical parameters, it is sufficient to mistakenly impose ``line equations" on the $\langle TTT\epsilon \rangle $ point degrees of freedom (i.e., introduce a handful of extra derivatives to our functional) to have the appearance of excluding the full parameter space of the Ising CFT\@.

\section{Implementation}
\label{sec:implementation}

\subsection{Generating SDPs}
\label{sec:building}

Semidefinite programs (SDPs) for bootstrapping correlators of $\s,\e,T$ are somewhat complex, and we needed to devote significant engineering effort to generating them efficiently. Two ideas are particularly important in our implementation:
\begin{itemize}
\item Parallelism. Recall that {\tt SDPB} scales well to several nodes and a large number of cores. (For example, each of our $\Lambda=43$ results were computed on 16 nodes with 128 cores each, for a total of 2048 cores.) In order to ensure that {\it generating\/} SDPs did not become a bottleneck, we needed to achieve similar scaling for computing conformal blocks and assembling them into {\tt SDPB} input. One way to improve parallelism is to, whenever possible, break computations into independent ``chunks" that can be handled separately on different cores. However, one must also be careful to control the memory usage of each chunk, since otherwise the number of simultaneous chunks could become memory limited, degrading overall performance.

\item Memoization.\footnote{This is not a typo, see \url{https://en.wikipedia.org/wiki/Memoization}.} We take care to recompute data only when necessary. For example, conformal blocks appearing in the four-point function of stress tensors $\<TTTT\>$ remain unchanged when parameters like $\De_\s$ and $\De_\e$ are varied. Thus, we pre-compute them and store them in a central location for re-use. As another example, during an OPE scan, only OPE coefficients of external operators are changing --- the remaining conformal blocks and crossing equations are unchanged, so we take care to only re-compute the contribution to an SDP from the external operator channel.
\end{itemize}

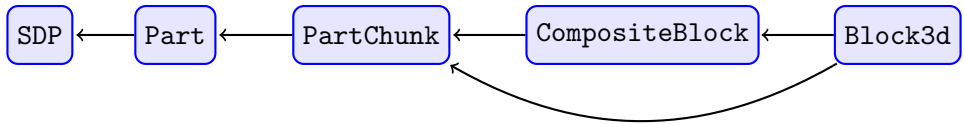
\begin{figure}
\centering
\begin{tikzpicture}[%
block/.style={
      rectangle,
      draw=blue,
      thick,
      fill=blue!10,
      rounded corners,
      minimum height=2em
    }
]
\node[block,right] at (0,0) {\tt SDP};
\node[block,right] at (1.7,0) {\tt Part};
\node[block,right] at (3.8,0) {\tt PartChunk};
\node[block,right] at (6.9,0) {\tt CompositeBlock};
\node[block,right] at (11,0) {\tt Block3d};
\draw[thick,->] (1.7,0) -- (0.93,0);
\draw[thick,->] (3.8,0) -- (2.83,0);
\draw[thick,->] (6.9,0) -- (5.94,0);
\draw[thick,->] (11,0) -- (10.04,0);
\draw[thick,->] (11.04,-0.38) to [out=-150,in=-30] (5.9,-0.4);
\end{tikzpicture}
\caption{\label{fig:sdpelements}Elements of an SDP\@. Arrows indicate reverse dependencies --- e.g.\ {\tt Block3d}'s are needed to compute a {\tt CompositeBlock}, while both {\tt Block3d}'s and {\tt CompositeBlock}'s are needed to compute a {\tt PartChunk}.}
\end{figure}

To achieve these aims of parallelism and memoization, we broke the process of building the SDP into several elements, illustrated in figure~\ref{fig:sdpelements}:
\begin{itemize}
\item {\tt Block3d}'s: These are the tables of conformal blocks computed by the program {\tt blocks\_3d}. Each run of {\tt blocks\_3d} is characterized by a unique {\tt BlockTableKey}, which includes information about the representations of the external operators, a choice of four-point structure, choices of three-point quantum numbers $j_{12},j_{43}$, together with settings like a choice of coordinates, recursion order, number of derivatives $\Lambda$, and precision. Different runs of {\tt blocks\_3d} can be performed in parallel. However, each run can require substantial computational resources. We describe our improvements to  {\tt blocks\_3d} to take advantage of multiple cores and reduce memory usage in section~\ref{sec:blocks3d}.

\item {\tt CompositeBlock}'s: A {\tt CompositeBlock} is a linear combination of {\tt Block3d}'s with coefficients that are rational functions of $\Delta$. Each {\tt CompositeBlock} is computed in a single-threaded Haskell process that reads the necessary {\tt Block3d} files, computes the appropriate linear combination, and saves the result to a file in binary format. Because these processes are single-threaded, we can run many of them at once. However, this requires keeping the memory usage of each process small. To achieve this, we implemented a streaming {\tt Block3d} parser that parses {\tt Block3d} data on the fly without reading whole files into memory. Also, when building linear combinations of blocks, we read a {\tt Block3d} from a file, add it to an accumulator  (with the appropriate coefficient), and then free its memory before proceeding to the next {\tt Block3d} file.\footnote{More precisely, we force evaluation of the accumulator with {\tt deepseq}, which ensures that unneeded data will be garbage collected.}

\item {\tt SDP PartChunk}'s: A {\tt PartChunk} is the contribution to a single crossing equation from a single channel. It is a symmetric matrix whose entries are vectors of derivatives of conformal blocks. Each {\tt PartChunk} can depend on several {\tt Block3d}'s and {\tt CompositeBlock}'s. Each {\tt PartChunk} is computed by a single-threaded Haskell process that reads the appropriate blocks, assembles them into a matrix, places them over a common denominator in $\Delta$ (if necessary), and writes the result to a binary file.

\item {\tt SDP Part}'s: A {\tt Part} is the contribution of a single channel to an SDP (it is either a positivity condition, a normalization condition for the SDP, or the objective for the SDP). It is a symmetric matrix whose entries are concatenated vectors of derivatives of conformal blocks. Each {\tt Part} is computed by a single-threaded Haskell process that reads the appropriate {\tt PartChunk}'s, concatenates their matrix entries, places the whole expression over a common denominator in $\Delta$, and writes the result to a JSON file in the appropriate format for {\tt pmp2sdp}.
\end{itemize}

Before building an SDP, we construct a tree of required elements and their dependencies.\footnote{We give details on how we determine these dependencies in appendix~\ref{app:dependencies}.} We then check whether any elements have already been built by checking for the existence of its corresponding file.\footnote{In order for this to be robust to crashes, it is important that the creation of a file be atomic. As an example of the type of situation we would like to avoid, suppose that a node crashes in the middle of saving a {\tt CompositeBlock} to a file, leading to a corrupted file. When we restart the computation, we would notice the existence of this file and conclude that we don't have to rebuild the corresponding {\tt CompositeBlock}. The whole program would then crash later when trying to incorporate the corrupted {\tt CompositeBlock} into a {\tt PartChunk}. To avoid this situation, we always write data to a temporary file, and then {\tt mv} the temporary file to its permanent location afterwards. The filesystem ensures that {\tt mv} is atomic.} If an element has already been built, we prune that part of the tree, removing the element and its dependencies. This achieves the goal of ``memoization" mentioned above. For example, if all of the elements of an SDP already exist, the tree becomes trivial. Finally, we convert the tree of elements to a tree of tasks. (For example, multiple {\tt Block3d's} corresponding to different spins {\tt j\_internal} can be combined into a single task --- i.e.\ a single run of {\tt blocks\_3d}.)

The tasks are distributed to available cores by a custom scheduler, which ensures that we stay within the limits of memory and available CPUs, while respecting the dependency order of tasks, and attempting to minimize the total time for building the SDP\@. (Since {\tt blocks\_3d} is multi-threaded, the scheduler must also decide how many cores to allocate to each {\tt blocks\_3d} process.) To make the scheduler as efficient as possible, we collect time and memory data for each task, and re-use those statistics when building subsequent SDPs. Our scheduler uses an ad-hoc greedy algorithm for distributing resources to tasks, described in appendix~\ref{app:scheduler}. This could be improved in the future by taking advantage of more sophisticated scheduling algorithms and/or open-source tools.

We give examples of the number of tasks needed to build an SDP, together with the total build time, in table~\ref{tab:numelts}.

\begin{table}
\centering
\begin{tabular}{l|c|c|c}
 & fresh build & using pre-built $\<TTTT\>$ blocks  & during OPE scan
\\
\hline
{\tt Block3d} & 629 & 79 & 0 \\
{\tt CompositeBlock} & 10538 & 4576 & 0 \\
{\tt PartChunk} & 7926 & 7926 & 53 \\
{\tt Part} & 363 & 363 & 1 \\
\hline
total time & 4.8 hours & 1.4 hours & 13 seconds
\end{tabular}
\caption{\label{tab:numelts}Number of each type of task needed to build an SDP with $\L=43$, together with the total build time on 16 Expanse nodes with 128 cores/node. (Note that a single {\tt Block3d} task produces multiple block tables, corresponding to different {\tt j\_internal}'s.) In a ``fresh" build (column 1), nothing is pre-built, and the total build time is 4.8 hours. During a fresh build, we store $\<TTTT\>$ blocks in a central location so that they can be re-used. A build making use of these pre-built $\<TTTT\>$ blocks has fewer {\tt Block3d}'s and {\tt CompositeBlock}'s to build, and takes 1.4 hours (column 2). Finally, during an OPE scan (after the initial SDP has already been built), we repeatedly change the OPE coefficients of external operators, but no other parts of the SDP change. Thus, no {\tt Block3d}'s or {\tt CompositeBlock}'s need to be re-built, and only the {\tt Part} corresponding to the external operators (and its corresponding {\tt PartChunk}'s) needs to be constructed again. The build time is negligible in this case.}
\end{table}

\subsection{Optimizing {\tt blocks\_3d}}
\label{sec:blocks3d}

A major contribution to SDP generation time comes from computing conformal blocks. We perform these computations using the program {\tt blocks\_3d}~\cite{Erramilli:2019njx,Erramilli:2020rlr}, which is capable of constructing numerical approximations to arbitrary conformal blocks in $d=3$~CFTs.

Prior to this work,~{\tt blocks\_3d} has been used in~\cite{Erramilli:2020rlr,Erramilli:2022kgp,He:2023ewx,Bartlett-Tisdall:2023ghh,Mitchell:2024hix} in bootstrap problems involving spinors and spin-1 currents, for which its performance proved to be adequate. In our setup, however, we found it necessary to implement significant improvements to the design of~{\tt blocks\_3d}. This was partly due to the need to frequently generate very complex conformal blocks (for instance, $\<\e TTT\>$ blocks need to be generated anew for every value of $\De_\e$), and partly due to the memory constraints of the clusters available to us.

In order to explain these improvements, we must give a brief overview of the function of~{\tt blocks\_3d}. A single run of {\tt blocks\_3d} computes multiple conformal blocks, which all share the same external operator representations, four-point structure label, as well as the $j_{12}$ and $j_{43}$ parts of the three-point function labels. The program outputs conformal blocks with all possible values of $j_{120}$ and $j_{430}$, where the index $0$ refers to the exchanged operator, as well as all values of the exchanged spin $j$ up to a cutoff $j_\text{max}$.

For each conformal block, an array of derivatives is returned, with each derivative approximated by a rational function of the exchanged scaling dimension $\De$. These approximations are computed using Zamolodchikov-like recursion relations derived in~\cite{Erramilli:2019njx}, see~\cite{Zamolodchikov:1987avt, Kos:2013tga,Kos:2014bka,Penedones:2015aga} for earlier work. A starting point for the recursion is the function $h_{\oo,j}(z,\bar z)$, given by a known analytic formula~\cite{Kravchuk:2017dzd, Erramilli:2019njx}, which controls the $\De\to \oo$ limit of the conformal block.

We denote the conformal block schematically by $g$. Prior to this work, the algorithm of~{\tt blocks\_3d} that computes $g$ could be described roughly as follows.
\begin{itemize}
	\item In the first stage, the derivatives $\ptl^m_r\ptl^n_\l g$ are computed.\footnote{Here, $r=\sqrt{\r\bar \r}$, $\l = \thalf\log(\r/\bar\r)$, and $\r=z/(1+\sqrt{1-z})^2$.} For this, the following steps are taken.
\begin{itemize}
	\item The derivatives $\ptl_r^m \ptl_\l^n h_{\oo,j}$ are evaluated.
	\item The derivatives $\ptl_r^m \ptl_\l^n h_{\oo,j}$ are used as a starting point for the recursive computation of the derivatives $\ptl^m_r \ptl^n_\l g$. The recursions for different values of $n$ are completely independent, and are performed in parallel.
	\item For some conformal blocks, only the $\ptl^m_r\ptl^0_\l g$ derivatives are required. In this case, only one recursion needs to be performed.
\end{itemize}
	\item In the second stage, the derivatives $\ptl^m_r\ptl^n_\l g$ are converted to the desired basis of derivatives ($x$,$t$-derivatives in our case).
\end{itemize}
Upon analyzing the performance of this algorithm as implemented in {\tt blocks\_3d}, we discovered several issues.
\begin{itemize}
	\item The evaluation of the derivatives $\ptl_r^m \ptl_\l^n h_{\oo,j}$ (the base case) was taking up the majority of the computation time.
	\item The parallelization by different values of $n$ had significant drawbacks. Firstly, the number of different values of $n$ for derivative order $\L$ is roughly $\L/2$, which is much smaller than the typical number of cores available on a compute node (compare $\L=43$ to a typical 128-core machine).
	\item The recursion part of the code used a significant amount of memory. Running $\L/2$ recursions simultaneously means that the memory consumption is multiplied by a factor of around 20 for $\L=43$. This meant that such high-$\L$ calculations could not fit into the memory of a single machine.
	\item An implementation error led to all derivatives $\ptl^m_r\ptl^n_\l g$ being computed in all cases, even when only $n=0$ was requested.\footnote{This was a minor performance loss in previous studies, but it makes a significant difference in our current study.}
\end{itemize}
To address these issues, we implemented the following optimizations.
\begin{itemize}
	\item We significantly optimized the code that evaluates $\ptl_r^m \ptl_\l^n h_{\oo,j}$ by adding memoization of some intermediate expressions (in addition to the memoization that was already in place).
	\item We revamped the structure of parallelism in~{\tt blocks\_3d}. Specifically, different values of $n$ are now processed sequentially, and we parallelize the recursion step instead. This leads to much better scalability and dramatically improves memory usage by avoiding memory-intensive tasks running in parallel. This new structure also allowed us to parallelize the computation of $h_{\oo,j}$, leading to further performance gains.
	\item As the restructuring required a significant refactoring of the codebase, we took the opportunity to implement many further minor optimizations. In particular, some memory-intensive memoizations have been removed without any negative impact on performance.
	\item We fixed the implementation errors in the case when only $n=0$ derivatives were requested.
\end{itemize}

These optimizations were implemented over a period of time of continuous development, and we were using the software the entire time. As a result, we have not performed a systematic comparison of the original version to the final version used in this project. However, in comparisons between intermediate versions we observed, for reasonably representative runs, a more than 10$\x$ improvement in memory usage and a similar improvement in computation (wall) time. Note that the improvements in memory usage also mean that more~{\tt blocks\_3d} instances can run on a single machine. This is parallelism with nearly perfect scaling, unlike the parallelization within a single {\tt blocks\_3d} instance,\footnote{This does not contradict the previously stated importance of this internal parallelization.} further reducing the time needed to generate all conformal blocks.

\subsection{Improvements to SDPB}

Our work benefited from major performance improvements in SDPB 3.0.\footnote{\url{https://github.com/davidsd/sdpb/releases/tag/3.0.0}} Here, we give a brief summary of the main changes. One of the most expensive global operations in each SDPB iteration is computing the matrix
$Q \equiv P^T P$, where
\be
P\equiv\left(\begin{array}{cccc}
	P_{1}\\
	P_{2}\\
	\vdots\\
	P_{J}
\end{array}\right)
\ee
is a dense tall matrix of multiprecision numbers, and each submatrix $P_i$ is calculated independently from a single positivity constraint. In this section, we refer to a positivity constraint as a ``block," since it usually comes from the contribution of a particular channel to a set of crossing equations.

\begin{figure}[]
	\centering{}
	\includegraphics[width=0.7\paperwidth]{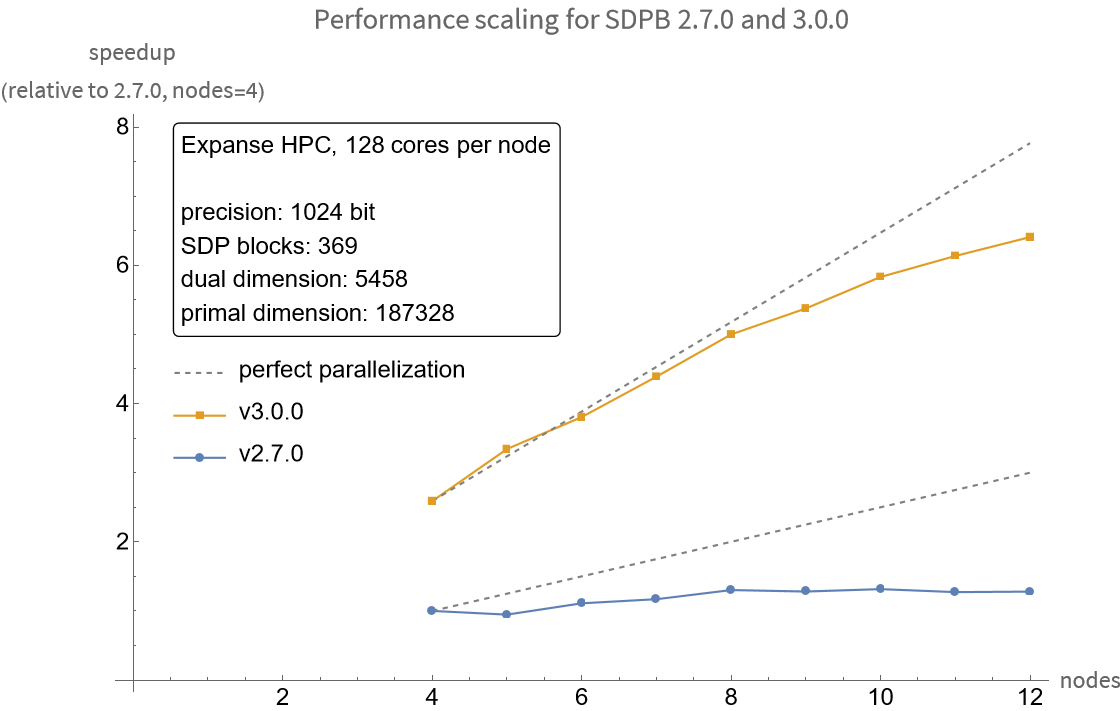}
	\caption{\label{fig:sdpb}SDPB 2.7.0 and SDPB 3.0.0 speedup on Expanse for the Ising model with stress tensors, $\Lambda=35$.
		The speedup is calculated relative to the smallest run for SDPB 2.7.0 (4 nodes).
		Each node has 128 cores.
}
\end{figure}

To compute $Q = P^T P$, one na\"ive approach is to compute the local contribution $Q_i \equiv P_i^T P_i$ from each block independently and then call reduce-scatter among cores to aggregate the results into the global distributed matrix $Q = \sum_i Q_i$. This algorithm was originally implemented in SDPB 2.0 and worked well for smaller problems.
However, it turned out to be a bottleneck as we proceeded to larger problems: reduce-scatter could take about 50\% of total execution time and thus ruined performance scaling for multi-node computations (see figure~\ref{fig:sdpb}).

In SDPB 3.0, two of us (VD and DSD) implemented a new matrix multiplication algorithm utilizing BLAS, the Chinese Remainder Theorem and MPI shared memory windows. Firstly, we now compute the contribution to $Q$ from the blocks on each node, and then perform reduce-scatter among nodes. This requires much less communication than reduce-scatter among cores and results in much better scaling.
Secondly, we made matrix multiplication on a node several times faster by utilizing highly efficient BLAS routines, which was not previously possible with multiprecision matrices.
To this aim, we split each multiprecision matrix $P'$ into several double-precision matrices $P'_{(k)}$, where $P'$ is the part of the matrix $P$ stored on a given node.
The matrices $P'_{(k)}$ are stored in a shared memory window, so that all cores on a node work together to calculate $Q'_{(k)} = P'^T_{(k)} P'_{(k)}$ using BLAS\@.
Then we restore the multiprecision matrix $Q' = P'^T P'$ from $Q'_{(k)}$ using the Chinese Remainder Theorem and call reduce-scatter to synchronize the result between nodes.

Figure~\ref{fig:sdpb} shows performance scaling of SDPB for a large-scale computation for the Ising model with stress tensors at $\Lambda=35$.
(This computation requires at least 4 Expanse nodes to fit in memory.) One can see that SDPB 3.0 is more than 2.5x faster than the previous version at 4 nodes, and scales well even for 10 nodes, achieving a factor of 6 speedup in that case.
A detailed description of the new matrix multiplication algorithm can be found at \url{https://github.com/davidsd/sdpb/blob/3.0.0/src/sdp_solve/SDP_Solver/run/bigint_syrk/Readme.md}.

\subsection{Approximating conformal blocks with interpolation}
\label{sec:interpolation}

In the course of this project we also found and implemented an improved approach to approximating conformal blocks using polynomials, leading to better-conditioned and more numerically stable SDPs. A commonly-used approach in the numerical bootstrap has been the following procedure \cite{Kos:2013tga}:
\begin{enumerate}
\item Truncate the radial coordinate expansion of a conformal block at some order, giving at the crossing-symmetric point an approximation $g_{\Delta} \approx r^{\Delta} \frac{P(\Delta)}{\Pi_p (\Delta - \Delta_p)}$ with a polynomial numerator $P(\Delta)$ and set of poles $\Delta_p$.
\item Choose a subset of poles according to a parameter $\kappa$, and perform ``pole-shifting," where we approximate $R(\Delta)=\frac{P(\Delta)}{\Pi_p (\Delta - \Delta_p)}$ in terms of a lower-degree rational function $S(\Delta)$ with only the subset of poles. This is usually done by imposing that some number of derivatives of $S(\Delta)$ around $\Delta=\Delta_\mathrm{min}$ and $\Delta=\oo$ should match the corresponding derivatives of $R(\Delta)$.
\end{enumerate}

We found that this procedure worked poorly in practice for our SDPs at high derivative order. The ``pole-shifting" approximation was not accurate enough. In order to make it more accurate, we had to significantly increase $\kappa$, but this led to larger degree polynomials, worse performance, and less numerical stability (forcing us to increase precision further, degrading performance even more). 

In our new approach, we approximate conformal blocks using interpolating polynomials, with a choice of sample points  specially adapted to the class of functionals we wish to describe. In practice, this leads to much more accurate approximations with much lower degree polynomials, improving both numerical stability and performance. We will describe the details of our approach in a separate publication~\cite{interpolation}.\footnote{The new interpolation scheme is already implemented and publicly available in the tool {\tt pmp2sdp} that is bundled with {\tt sdpb}.}

\subsection{Navigator and skydiving}
\label{sec:navigator}

The navigator function \cite{Reehorst:2021ykw} and skydiving algorithm \cite{Liu:2023elz} are approaches to optimizing SDPs that depend on parameters $\vec x$ (e.g.\ dimensions and OPE coefficients of ``external" operators). The navigator function $\cN(\vec x)$ is positive for disallowed points and negative for allowed points. Thus, one can then use gradient information about $\cN(\vec x)$ to guide the search for allowed points. The skydiving algorithm is a way to simultaneously solve for $\cN(\vec x)$ while moving in $\vec x$-space, potentially allowing one to find allowed points with a single {\tt sdpb} run. Both ideas can potentially significantly reduce the number of {\tt sdpb} iterations needed to map out an allowed space. However, a tradeoff is that one must typically build a very large number of conformal blocks and SDPs, since the external parameters $\vec x$ change many times.

In this work, we explored using skydiving with the generalized free field (GFF) navigator of \cite{Reehorst:2021ykw}.
Our external parameters are:
\be\label{eq:x}
\vec x = \left(\De_\s,\De_\e,  \frac{\l_B}{\l_{TT\e}}, \frac{\l_F}{\l_{TT\e}}, \frac{\l_{\s\s\e}}{\l_{TT\e}}, \frac{\l_{\e\e\e}}{\l_{TT\e}}\right).
\ee
We ran the skydiving algorithm with three different objectives:
\begin{enumerate}
\item Minimize $\mathcal N(\vec x)$ to find a point deep in a bounded feasible region.
\item Maximize $\De_\s$ to bound the island from the right.
\item Minimize $\De_\s$ to bound the island from the left.
\end{enumerate}
In practice, we found that skydiving was effective at lower derivative orders $\Lambda=11,19$ (and partially at $\Lambda=27,35$). However, at higher derivative orders, the cost of generating many SDPs became prohibitive. It was more efficient to spend computational time in {\tt sdpb} than in generating many new SDPs. Moreover, the gradient descent cannot be easily parallelized. Hence, at higher derivative orders we preferred the ``cutting surface" algorithm and Delaunay triangulation approach of~\cite{Chester:2019ifh}.

\subsection{Code availability and ancillary files}

Our code is open-source and available online. This includes the programs listed in appendix A of \cite{Erramilli:2022kgp}, as well as the central repository for this work:
\begin{itemize}
\item \url{https://gitlab.com/davidsd/stress-tensors-3d} \\
commit hash: {\tt 87d8076ee36373848d5cc3f4f5ac9c99856b9ac3}\\
 This repository includes setup code for constructing SDPs from $T,\sigma,\epsilon$ three-point structures and crossing equations, an implementation of the ``scheduler" described in appendix~\ref{app:scheduler}, versions of the OPE cutting surface and skydiving algorithms specialized to use the scheduler, as well as all the programs used to generate bounds for this work.
\end{itemize}
In addition, we include the following ancillary files with this paper:
\begin{itemize}
\item {\tt three\_point\_structures.m}: The three-point structures used in this work, in {\it Mathematica\/} format.
\item {\tt structures\_3d.m}: A {\it Mathematica\/} package for manipulating three-point structures, including performing conversions between embedding-space structures, the $q$-basis, the $\SO(3)_r$ basis, and more.
\end{itemize}

\section{OPE Scan and certifying disallowed points}
\label{sec:certifying}

\subsection{Review: cutting surface algorithm}

The crossing equations for $T,\s,\e$ system, when expanded in conformal blocks, contain contributions from the external operators $T,\s,\e$ themselves. Schematically, this can be written as
\be\label{eq:crossing_single_contrib}
	\vec \l^T \cG_\text{ext}^I \vec \l + \dots =0,
\ee
where
\be
	\vec \l = (\l_B,\l_F,\l_{\s\s\e},\l_{\e\e\e},\l_{TT\e})^T
\ee
is the vector of OPE coefficients involving the external operators and $\cG_\text{ext}^I$ is a symmetric $5\x 5$ matrix constructed out of derivatives of conformal blocks, and $I$ labels different components of the crossing equations.

The basic approach in numerical bootstrap involves searching for a linear functional $\a_I$ such that, among other constraints, the matrix $\a_I\cG_\text{ext}^I$ is positive semi-definite,
\be\label{eq:naivePSD}
	\a_I\cG_\text{ext}^I\succeq 0.
\ee
Acting with such an $\a_I$ on the crossing equations we get
\be
	\vec \l^T (\a_I\cG_\text{ext}^I) \vec \l + \dots =0.
\ee
The constraint~\eqref{eq:naivePSD} ensures that the first term is non-negative. The other constraints on $\a_I$ ensure that all the remaining terms are non-negative and that there is at least one positive contribution, leading to a contradiction. This idea allows us to exclude the values in the parameter space for which a suitable $\a_I$ can be found.

It has been observed in~\cite{rychkov_unpublished, Kos:2016ysd} that the constraint~\eqref{eq:naivePSD} is in fact too strong. It also excludes solutions to crossing equations where there are several distinct contributions of the same form in the crossing equations,
\be
	\sum_i \vec \l^T_i \cG_\text{ext}^I \vec \l _i+ \dots =0.
\ee
However, we know that in Ising CFT there are unique operators $\s,\e,T$ with the respective quantum numbers, leading to single term as in~\eqref{eq:crossing_single_contrib}. 

A better strategy is to treat the vector $\vec\l$ as a parameter similar to the scaling dimensions $\De_\s,\De_\e$, and to try to exclude a given value of $\vec\l$ by searching for $\a_I$ that instead of~\eqref{eq:naivePSD} satisfies
\be
	\vec \l^T (\a_I\cG_\text{ext}^I) \vec \l\geq 0.
\ee
Importantly, this constraint is unchanged if we rescale $\vec\l$ by a non-zero real number. Therefore, it only depends on the equivalence class $[\vec\l]$ of $\vec\l$ in $\R\P^4$ and we should treat $[\vec\l]$ as the new parameter. If we are able to exclude all possible values of $[\vec \l]$ for a fixed $(\De_\s,\De_\e)$, then the point $(\De_\s,\De_\e)$ is excluded in the dimension space. As shown in~\cite{Kos:2016ysd}, this procedure yields a significantly smaller allowed region for $(\De_\s,\De_\e)$ than using~\eqref{eq:naivePSD}.

How can one exclude all possible values of $[\vec \l]$ in a finite amount of time? The approach used in~\cite{Kos:2016ysd} was to discretize $\R\P^1$ (their $\vec\l$ had only two components) using a finite grid. If all points in the grid are excluded, one can be reasonably confident that there are no allowed points in $\R\P^1$ if the grid is fine enough. This approach worked well in~\cite{Kos:2013tga} because their parameter space $\R\P^1$ was one-dimensional. In our case the parameter space is $\R\P^4$ and discretization appears computationally infeasible.

This problem was solved in~\cite{Chester:2019ifh} with the ``cutting surface'' algorithm. The key observation is simply that a finite set of functionals $\a_I^{(k)}$, $k=\{1,2,\dots, N\}$, can be enough to exclude the full parameter space $\R\P^4$. Indeed, existence of a functional $\a_I^{(k)}$ implies that all $[\vec \l]$ which satisfy
\be
	\vec\l^T(\a_I^{(k)}\cG_\text{ext})\vec\l\geq 0
\ee
are excluded. This condition defines a closed set $E_k$, and it might very well be that the union of all $E_k$ covers $\R\P^4$,
\be\label{eq:allexcluded}
	\bigcup_{k=1}^N E_k = \R\P^4.
\ee

If~\eqref{eq:allexcluded} holds, then all $[\vec\l]$ are excluded. If~\eqref{eq:allexcluded} doesn't hold, then we can take a $[\vec\l_{N+1}]$ that is not in $\bigcup_{k=1}^N E_k$, i.e.
\be\label{eq:newlambdacondition}
	\vec\l_{N+1}^T(\a_I^{(k)}\cG_\text{ext})\vec\l_{N+1}< 0,\quad \forall k\in\{1,2,\dots,N\},
\ee
and try to construct a linear functional $\vec\a^{(N+1)}_I$ such that (along with all the other positivity constraints for other contribution to the crossing equations)
\be
		\vec\l_{N+1}^T(\a_I^{(N+1)}\cG_\text{ext})\vec\l_{N+1}\geq 0.
\ee
If such a functional $\a_I^{(N+1)}$ is found, we can add it to the list, increment $N$, and check~\eqref{eq:allexcluded} again. If a functional cannot be found, it means that $[\vec\l_{N+1}]$ is allowed and so are the current values of $(\De_\s,\De_\e)$. Typically, this process converges in a number of steps which is much smaller than the number of points in a fine enough grid on $\R\P^4$. It also has the benefit of not relying on a discretization.

The technical content of the cutting surface algorithm~\cite{Chester:2019ifh} is in the procedure for finding a solution $[\vec\l_{N+1}]$ to~\eqref{eq:newlambdacondition} or determining that no solution exists.\footnote{When several solutions to~\eqref{eq:newlambdacondition} can be found, the cutting surface algorithm will try to select a solution that is the furthest away from the sets $E_k$ in order to speed up the convergence of the whole process.} Abstractly, this can be formulated as the following problem: given quadratic forms $Q_0, Q_1,\dots, Q_N$, determine whether there exists a solution $\vec \l\in \R^{n+1}$ to the equations
\be\label{eq:QCQP}
	\vec \l^T Q_k \vec \l <0,\quad k\in \{0, 1,2,\dots,N\}.
\ee
This is a quadratically-constrained quadratic problem; this class of problems is in general NP-hard. For this reason, the cutting-surface algorithm of~\cite{Chester:2019ifh} uses a number of methods to try to solve~\eqref{eq:QCQP}. If no solution is found by these methods, the algorithm proceeds assuming that no solution exists, but this decision is based on heuristics and is not rigorous.

In this work we developed an efficient algorithm that can rigorously certify that no solutions to~\eqref{eq:QCQP} exist. This does not contradict NP-hardness of the problem since our algorithm is not guaranteed to construct a certificate. However, in all cases when the cutting-surface algorithm of~\cite{Chester:2019ifh} terminated with ``no solution found'', our algorithm was able to certify that no solutions exist. Therefore, all disallowed points in our plots have been ruled out rigorously.

Before we describe the algorithm, we note that in practice while the quadratic forms $Q_1,\dots, Q_N$ are given by the functionals $\a_I^{(k)}$ as
\be
	Q_k = \a_I^{(k)}\cG^I_\text{ext},
\ee
the quadratic form $Q_0$ is specified manually to restrict the search to an ellipse in $\R\P^{n+1}$. This bounding ellipse is chosen conservatively to include the region where the allowed points $[\vec\l]$ are expected to be. See section~\ref{sec:mixedresults} for details. In essence, one of our assumptions is that if any $[\vec \l]$ is allowed at all, it is deep inside the region defined by $\vec\l^T Q_0\vec\l<0$. The technical meaning of this will be clarified below.

\subsection{Certification algorithm}

In what follows we will use the notation $Q(\l)=\l^T Q \l$ for quadratic forms. Since $Q_0$ is specified by us and is chosen so that $Q_0(\l)<0$ represents an ellipsoid, it has signature $(1,n)$. By a change of basis, we can assume that 
\be
	Q_0(\l) = -t^2+y^2,
\ee
where $\l=(t,y)$ with $t\in \R$ and $y\in \R^n$.

Recall that we are looking for a solution $\l\in \R^{n+1}$ to 
\be
	Q_k(\l)<0,\quad k=0,1,\dots N.
\ee
Any solution satisfies $Q_0(\l)=-t^2+y^2<0$. Therefore, $t\neq 0$ and by rescaling $x$ we can always set $t=1$. Therefore, it is enough to find a solution $y\in \R^n$ to
\be\label{eq:yequations}
	Q_k((1,y))<0,\quad k=1,\dots, N,
\ee
such that $|y|<1$ or prove that none exists.

We can write
\be
	Q_k((1,y)) = c_k+\b_k(y)+w_k(y),
\ee
where $c_k\in \R$, $\b_k$ is a linear functional and $w_k$ is a quadratic form. It was observed in~\cite{Chester:2019ifh} that in bootstrap applications the surfaces $Q_k((1,y))=0$ are usually approximately flat inside the ball $|y|<1$. In other words, the coefficients in the quadratic form $w_k$ are much smaller than $c_k$ and $b_k$, and for $|y|<1$ the contribution of $w_k(y)$ to $Q_k((1,y))$ is subleading. Neglecting $w_k(y)$ would be useful since that would turn~\eqref{eq:yequations} into linear inequalities.

Since our goal is to construct a rigorous proof that no solutions to~\eqref{eq:yequations} exist, we cannot simply neglect $w_k(y)$. Instead, we will search for a rigorous linear lower bound on $Q_k$ of the form
\be\label{eq:Qbound}
	Q_k((1,y)) \geq c_k+\de c_k+\b_k(y).
\ee
Let 
\be
	\cA =\{y\in \R^n| |y|<1 \text{  and  }Q_k((1,y))<0, \quad k=1,\dots,N\}
\ee
be the set of solutions to~\eqref{eq:yequations} (our goal is to show that $\cA=\emptyset$). Suppose we can show that
\be\label{eq:dck_constraint}
	w_k(y)\geq \de c_k,\quad \forall y\in\cA.
\ee
Then~\eqref{eq:Qbound} holds on $\cA$ and therefore
\be
	\cA\subseteq \cB[\de c]=\{y\in \R^n| c_k+\de c_k+\b_k(y)<0, \quad k=1,\dots,N\}.
\ee
The set $\cB[\de c]$ is defined using linear inequalities, and therefore can be studied using linear programming. In particular, it is easy to check whether $\cB[\de c]=\emptyset$. If we can show that $\cB[\de c]=\emptyset$,  this would constitute a proof of $\cA=\emptyset$. Our strategy will therefore be to construct various sets of coefficients $\de c_k$ for which~\eqref{eq:dck_constraint} holds. Our notation $\cB[\de c]$ stresses the fact that $\cB[\de c]$ depends on the choice of the $\de c_k$.

A simple choice is to set 
\be
	\de c_k = \de c_k^{0}\equiv  \inf_{|y|<1} w_k(y) = \min\{\l_{k,\text{min}},0\},
\ee
where $\l_{k,\text{min}}$ the minimal eigenvalue of the matrix that defines $w_k$. With this choice~\eqref{eq:dck_constraint} is obviously true, and therefore $\cA\subseteq \cB[\de c^0]$. Typically, we find that the set $\cB[\de c^0]$ is not empty. It is, however, usually much smaller than the unit ball $|y|<1$. This allows us to improve our choice of $\de c_k$.

We construct an improved set of constants $\de c_k^1$ as follows. Let $\l_{k,i}$ and $e_{k,i}$ be the eigenvalues and the eigenvectors of $w_k$. We can then write
\be
	w_k(y)& = \sum_{i}\l_{k,i} (e_{k,i}\. y)^2\geq \de c_k^1,\\
	\de c_k^1&\equiv \sum_{i:\l_i>0}\l_{k,i}\inf_y (e_{k,i}\.y)^2+\sum_{i:\l_i<0}\l_{k,i}\sup_y (e_{k,i}\.y)^2.
\ee
We would like this to hold on $\cA$. Since $\cA\subseteq\cB[\de c^0]$, this can be achieved by taking the $\sup$ and the $\inf$ over $\cB[\de c^0]$. Note that $(e_{k,i}\.y)$ is a linear function on the convex set $\cB[\de c^0]$, so its image is an interval which can be computed using linear programming. From this, $\inf_{y\in \cB[\de c^0]} (e_{k,i}\.y)^2$ and $\sup_{y\in \cB[\de c^0]} (e_{k,i}\.y)^2$ can be easily obtained.\footnote{We also implemented two minor improvements on the value of $\de c^1$. Firstly, it is guaranteed that $(e_{k,i}\.y)^2\leq 1$ on $\cA$, so we can replace $\sup_{y\in \cB[\de c^0]} (e_{k,i}\.y)^2$ by $\min\{\sup_{y\in \cB[\de c^0]} (e_{k,i}\.y)^2, 1\}$. Secondly, one can replace $\sum_{i:\l_i<0}\l_{k,i}\sup_y (e_{k,i}\.y)^2$ by $\max\{\sum_{i:\l_i<0}\l_{k,i}\sup_y (e_{k,i}\.y)^2, \de c^0_k\}$.}

In the majority of cases when the cutting surface algorithm of~\cite{Chester:2019ifh} does not find a solution, we observe that the set $\cB[\de c^1]$ is empty. Since $\cA\subseteq \cB[\de c^1]$, this proves that $\cA=\emptyset$ and certifies the absence of solutions. When the set $\cB[\de c^1]$ is non-empty, we can construct $\de c^2$ from $\cB[\de c^1]$ just as $\de c^1$ is constructed from $\cB[\de c^0]$. If $\cB[\de c^2]$ is non-empty, we can continue to construct $\de c^3$ and so on. With rare exceptions discussed below, we found in practice that this process terminates with an empty set $\cB[\de c^m]$ for some $m>0$. The largest value of $m$ that we observed was $m=3$. 

Out of the 38 disallowed points in $(\De_\s,\De_\e)$ space that we computed at $\L=43$, 30 points were certified by $\cB[\de c^1]=\emptyset$, 1 point by $\cB[\de c^2]=\emptyset$, and 2 points by $\cB[\de c^3]=\emptyset$. This produces certificates for 33 out of 38 points.

In these cases the sets $\cB[\de c^m]$ are typically quite small and deep in the interior of the unit ball $|y|<1$. However, for the 5 remaining points $\cB[\de c^0]$ is an unbounded set, and $\de c^1$ cannot be constructed. It appears that this happens when the cutting surface algorithm quickly excluded the majority of the unit ball $|y|<1$, and then spent some time ruling out a boundary region close to $|y|=1$.

To deal with such situations, we use convex quadratic optimization\footnote{We used the function {\tt QuadraticOptimization} in {\tt Mathematica}.} to compute
\be
	r_\text{min}\equiv \inf_{y\in \cB[\de c^0]}|y|.
\ee
We found that in all 5 cases at $\L=43$ when $\cB[\de c^m]$ iterations fail, $r_\text{min}$ is at least 0.99. Since $\cA\subseteq \cB[\de c^0]$, this shows that any solution to~\eqref{eq:yequations} satisfies $|y|>0.99$.

Recall that the ball $|y|<1$ is the bounding ellipse that we choose conservatively to restrict the search region in the OPE space. In other words, we are assuming that all solutions, if any, lie deep inside the unit ball $|y|<1$. In particular, we assume that no solution can satisfy $|y|>0.99$. Therefore, the statement that all solutions satisfy $|y|>0.99$ is a proof that no solutions exist under our assumptions.

\section{Results}
\label{sec:results}

\subsection{The $\<TTTT\>$ system}
\label{sec:ttttresults}

We first present new results in the system of only correlators of \(T\), i.e.~constraints on \(\<TTTT\>\). This was previously studied in~\cite{Dymarsky:2017yzx}, where the authors found universal bounds for all local unitary 3d CFTs on the spectrum of low-lying operators as well as the surprising presence of features seemingly linked to the 3d Ising CFT in particular. As we have a completely new implementation of this system (which is a subset of our full \(T,\s,\e\) system) and have the target of the 3d Ising CFT, it's therefore worthwhile to dwell briefly on this \(T\)-only system. The results serve to validate the correctness of our implementation, to improve and refine the universal numerical results of~\cite{Dymarsky:2017yzx}, and to reveal what more these universal bounds seem to know about the 3d Ising CFT\@.

\subsubsection{Bounds on scalar gaps}

\begin{figure}
    \hspace{-1.3cm}
    \centering
    \includegraphics[width=15cm]{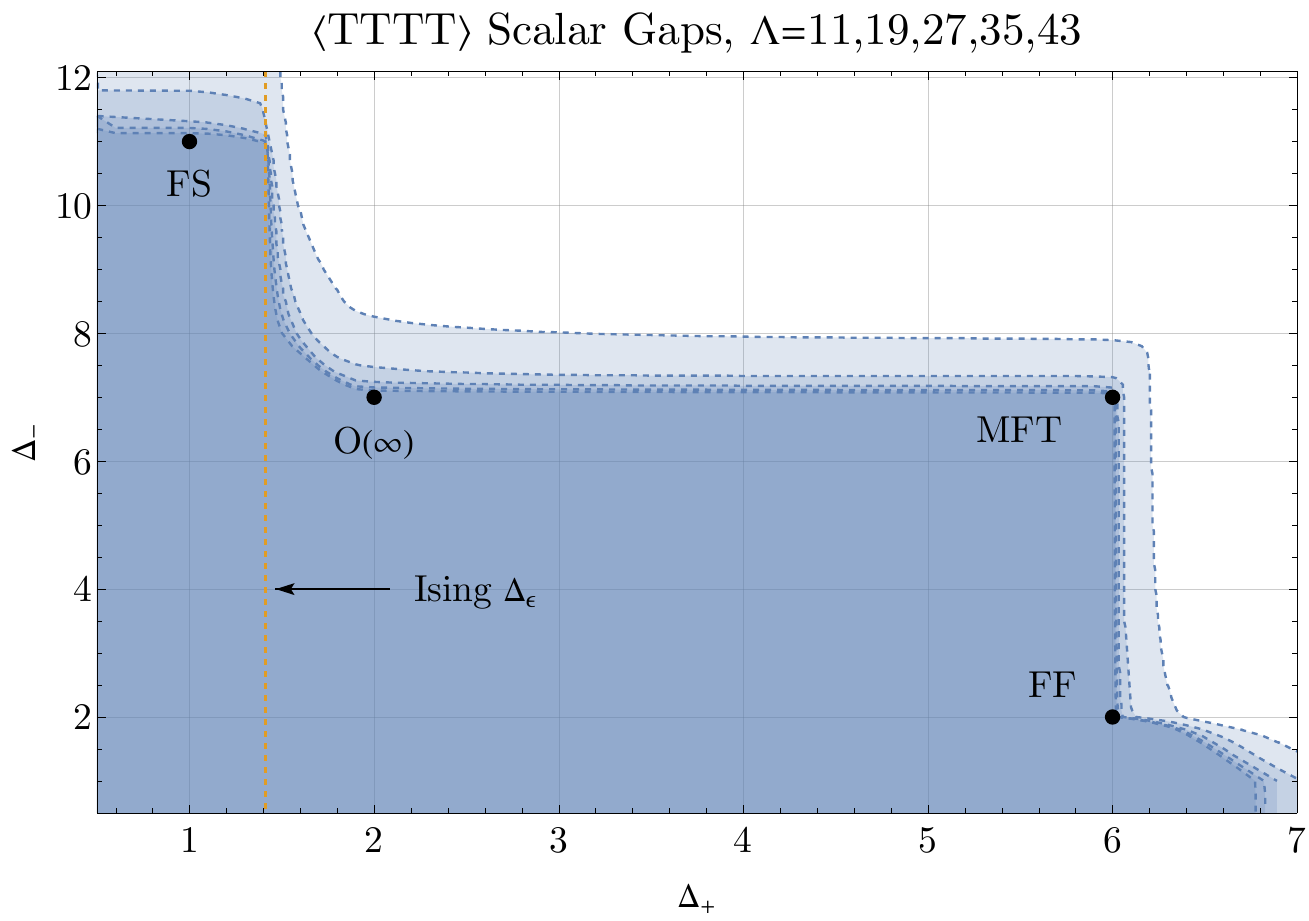}
    \caption[\(\<TTTT\>\) Scalar Gaps]{\label{fig:tttt_scalars} Bound on the allowed gaps for parity-even and -odd scalars in the single-correlator \(\<TTTT\>\) system. Successively interior regions correspond to higher derivative order \(\Lambda\). We can see that by \(\Lambda=43\) the bounds are fairly well converged and therefore are hard to distinguish visually at this scale. The known values for these gaps in the free scalar (FS) theory, Majorana free fermion (FF) theory, stress tensor mean-field theory (MFT), \(O(N)\) CFT in the large-\(N\) limit (\(O(\infty)\)), and Ising CFT are indicated.}
\end{figure}

\begin{figure}
    \hspace{-1.3cm}
    \centering
    \includegraphics[width=12cm]{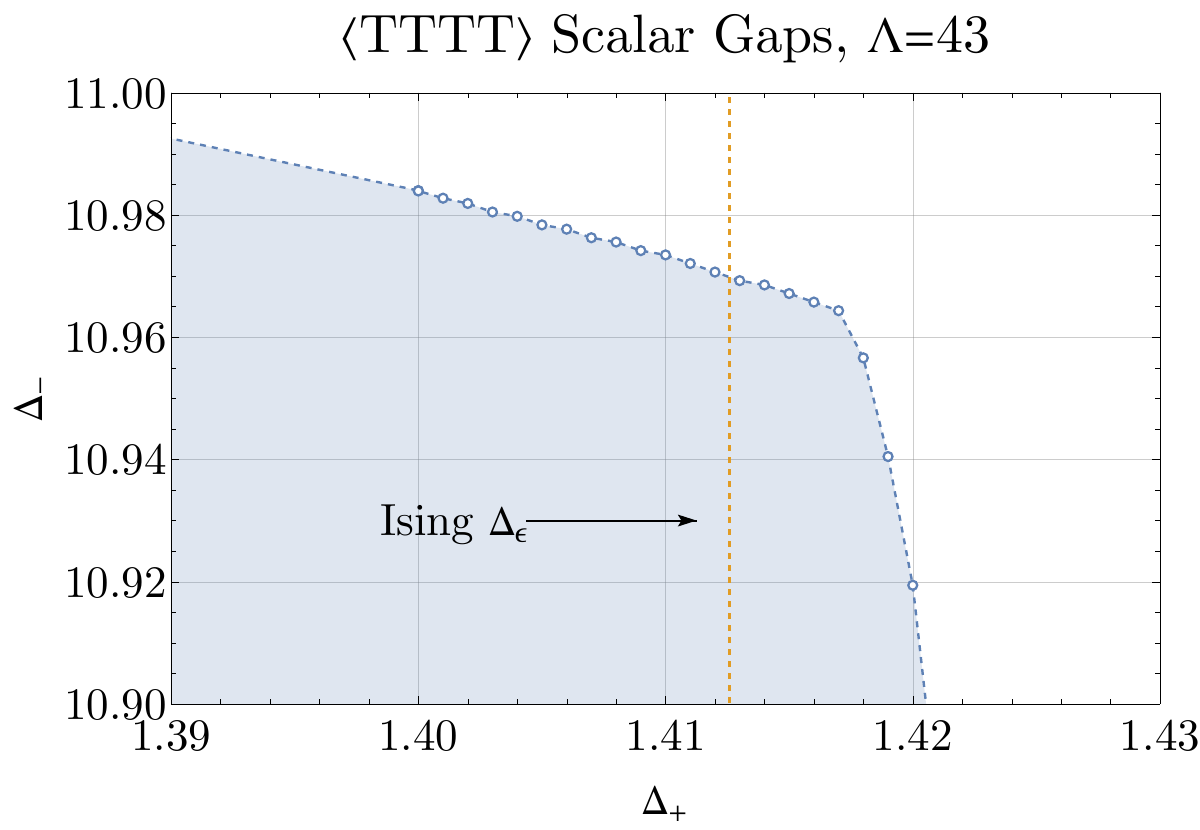}
    \caption[\(\<TTTT\>\) Scalar Gaps Ising Kink]{\label{fig:tttt_scalars_ising} Inset around the Ising CFT data of the \(\Lambda=43\) bound on the allowed gaps for parity-even and -odd scalars in the single-correlator \(\<TTTT\>\) system; see figure \ref{fig:tttt_scalars} for the full bound. The known value for the \(\Delta_+\) gap in the Ising CFT is indicated. The markers indicate data points.}
\end{figure}

The headline result of~\cite{Dymarsky:2017yzx} was the discovery of universal upper bounds on the dimensions of the leading scalars, both parity even and odd, for local unitary 3d CFTs. In particular,~\cite{Dymarsky:2017yzx} found a contour in the space of \((\Delta_+,\Delta_-)\) (\(\pm\) referring to the respective space parities) up to the derivative order of $\Lambda = 19$. We reproduce and improve on this result up to $\Lambda=43$ in figure \ref{fig:tttt_scalars}. This contour features sharp discontinuities at various notable CFTs such as the free Majorana fermion CFT and the Ising CFT, as well as the stress tensor mean-field theory. We refer to~\cite{Dymarsky:2017yzx} for further details.

We obtained the result in figure~\ref{fig:tttt_scalars} with two grids of binary searches, namely by fixing \(\Delta_+\) while searching over \(\Delta_-\) and vice versa. The fixed values sample a range from \(\frac{1}{2}+10^{-6}\) (just above the unitarity bound) to where there are no longer any allowed points. The only gap assumptions for this bound are $\De_+$ and $\De_-$ in parity-even and parity-odd scalar channels. 

We see that our results both agree with and improve upon the prior result of~\cite{Dymarsky:2017yzx}. Notably, the discontinuity near the Ising CFT persists to $\Lambda=43$, and by fixing \(\Delta_+=1.41262528\), i.e.\ our central value of $\De_\e$ in the Ising CFT, we find the upper bound \(\Delta_-\lesssim 10.97\). We show a zoomed view of this bound in figure \ref{fig:tttt_scalars_ising}. This is close to expectations based on the free scalar value 11, and it's tempting to speculate that the Ising CFT nearly saturates this universal bound on $\De_-$. We will have more to say in the next section.

In~\cite{Dymarsky:2017yzx} the authors also noted an intriguing feature that appears at \((\Delta_+,\Delta_-)=(7,1)\) which seemed to not be any known CFT\@. Even at $\Lambda=27$, which is our first step beyond the previous work's numerical order, the bound gets stronger and leaves \((7,1)\) disallowed. We therefore conclude that this neighborhood of \(\Delta_+>6\) is likely to not contain any physical CFTs and that the presence of the feature in the bound likely amounts to some low-derivative order numerical artifact.\footnote{Indeed, the authors of the previous work note that this region is disallowed if one assumes that the same leading parity even scalar also appears in the OPE of \(\mathcal{O}_-\times\mathcal{O}_-\), i.e.\ the leading odd scalar with itself. So for it to be disallowed even without such an inoffensive assumption would lend credence to this region not having much physical interest.}

\subsubsection{Bounds on OPE coefficients}

\begin{figure}
    \hspace{-0.75cm}
    \centering
    \includegraphics[width=16cm]{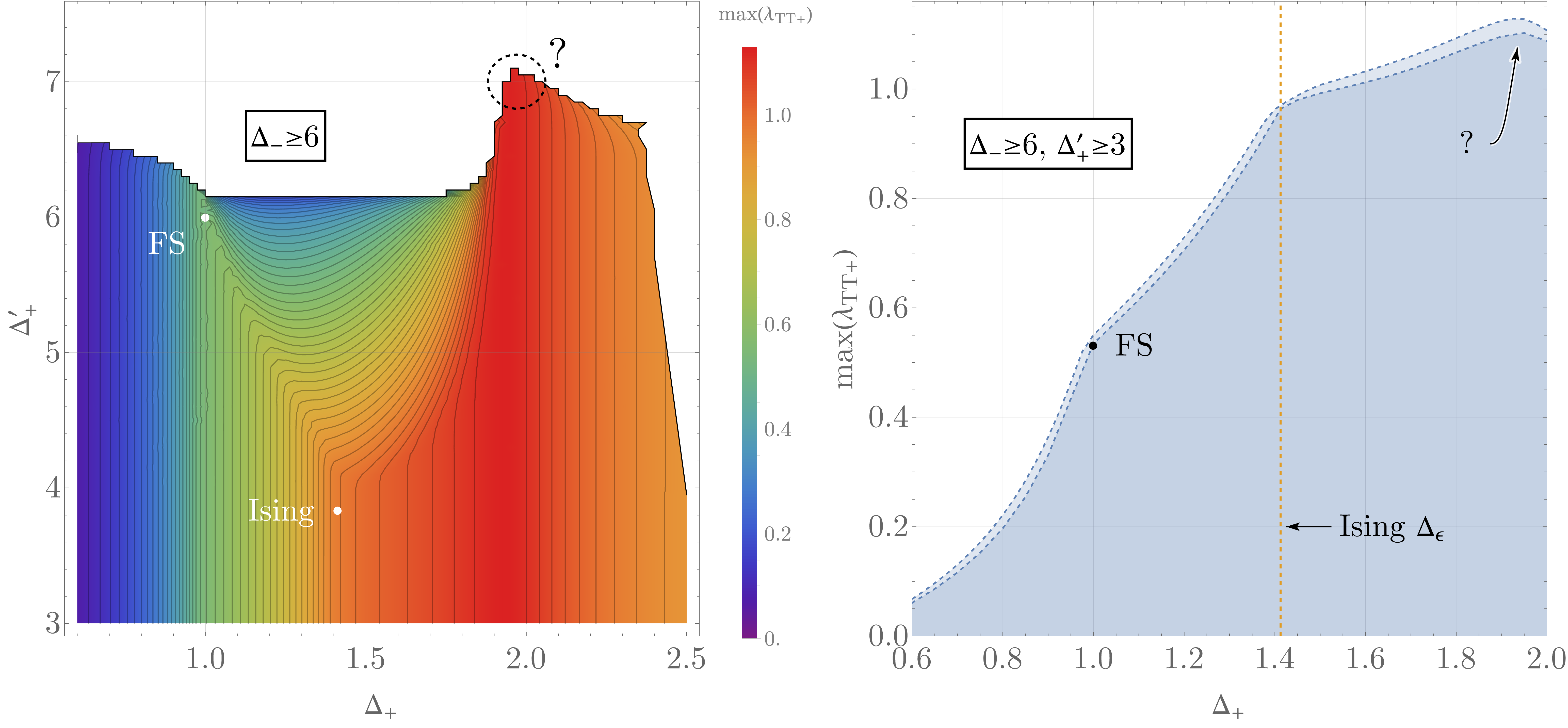}
    \caption[\(\< TTTT\>\) OPE maximization for \(\lambda_{TT+}\)]{\label{fig:tte_bound} Results from OPE maximization of the coefficient \(\lambda_{TT+}\), corresponding to the leading parity-even scalar, as a function of its dimension \(\Delta_+\) and the gap to the next parity-even scalar \(\Delta'_+\), subject to the assumption that the leading parity-odd scalar has dimension \(\Delta_-\geq 6\). The left-hand plot presents the results for \(\Lambda=19\) as a contour plot, and the right-hand plot presents the results for \(\Lambda=19,27\) as a line plot for the slice of \(\Delta'_+\geq 3\). In both, we highlight the known solution of the free scalar (FS) theory,\textsuperscript{\ref{fn:free}} the best-known numerical bootstrap results for the Ising CFT, and a third mysterious feature. }
\end{figure}

Given the presence and persistence of an Ising kink in these universal bounds of local unitary 3d CFTs, one should wonder how much more the pure \(\<TTTT\>\) system can tell us about the Ising CFT\@. For example, can we see the Ising CFT in bounds of OPE coefficients? Is Ising extremal in the space of CFTs, and how much so? To answer this, we computed upper bounds on the OPE coefficient of the leading even scalar \(\lambda_{TT+}\),\footnote{See equation~\eqref{eq:tte} for the explicit statement of the conventions.} corresponding to \(\lambda_{TT\e}\) for the Ising CFT\@. This coefficient was recently studied non-rigorously by both the fuzzy sphere~\cite{Hu:2023xak} and the 5-point bootstrap~\cite{Poland:2023vpn, Poland:2023bny}, but to our knowledge had not been studied rigorously by the 4-point bootstrap before.

We computed the upper bound on \(\lambda_{TT+}\) as a function of \(\Delta_+\) and \(\Delta'_+\), where \(\Delta'_+\) is the gap to the next-to-leading parity-even scalar. We made a conservatively Ising-compatible assumption of \(\Delta_- \geq6\).  Our results are presented in figure \ref{fig:tte_bound}.

Let us focus on the slice with the assumption \(\Delta'_+\geq3\). Remarkably, we see that even with these few assumptions the upper bounds on \(\lambda_{TT+}\) show sharp and distinct features. We first see a discontinuity at \(\Delta_+=1\) which exactly corresponds to the free scalar theory including the known value of the OPE coefficient; so the free scalar saturates these bounds.\footnote{\label{fn:free}We note that this kink is present as a vertical line on the full \((\Delta_+,\Delta'_+)\) plot all the way to a gap assumption of \(\Delta'_+>6\). The careful reader may worry that this seems to be inconsistent with the free scalar theory which has an operator \(\phi^4\) of dimension 2. However, such an operator does not and cannot appear in the OPE of \(T\times T\); equivalently \(\langle T T \phi^4\rangle=0\). Indeed, Wick's theorem factorizes \(\langle T T \phi^4\rangle=0\) into two point functions \(\langle T \phi^2\rangle^2\), which then must vanish by conformal symmetry. The leading scalar that appears in this OPE is therefore of the form \(T^2\), of dimension 6, which is exactly in line with our bound.}  Then, for \(\Delta_+\sim1.41\) we have another sharp feature, evidently corresponding to the Ising CFT\@. The upper bound at this $\De_+$ is consistent with previous estimates of $\l_{TT\e}$ in the Ising CFT in~\cite{Hu:2023xak,Poland:2023vpn, Poland:2023bny}. As we will see in the next section, the bound is in fact nearly saturated by the high-precision value of $\l_{TT\e}$ in the Ising CFT!

We can see more evidence to this effect even without the exact value of this OPE coefficient. Returning to the full \((\Delta_+,\Delta'_+)\) plot, note that this putative Ising OPE kink disappears roughly around \(\Delta'_+\sim 3.9\), which is just above where the known Ising CFT would lie, with \(\Delta_{\epsilon '}=3.82951(\mathbf{61})\) \cite{Reehorst:2021hmp}. This further underlines that these features correspond to the Ising CFT, and that it appears to be extremal --- that is, special enough to be found even without assumptions of symmetries in this universal system of purely stress tensors, in multiple ways. One might wonder if this is a feature of the Ising CFT itself, or if there are other CFTs which are similarly extremal.

On this topic, we note that there appears to be a third such feature in the OPE coefficient bound, denoted by ``?''. It appears to correspond to roughly \((\Delta_+,\Delta'_+)\sim (1.9,7)\), which isn't obviously related to any known CFT\@. Some guesses are that it is related to the \(\mathcal{N}=2\) super-Ising CFT, an $N=4$ Gross-Neveu-Yukawa CFT, or an \(O(N)\) model at some large value of $N$, which have $\Delta_+$ near \(1.9\)~\cite{Bobev:2015vsa, Erramilli:2022kgp}; however these theories do not appear consistent with the $\Delta'_+$ gap of 7~\cite{Henriksson:2022rnm, Mitchell:2024hix}. It could also be possible that this feature is a numerical effect which lacks any physical correspondence. Regardless, this mysterious feature warrants further investigation.

\subsection{Mixed correlators of $T,\s,\e$}
\label{sec:mixedresults}

Now we will move on to the main focus of this work, which is to study the system of $T,\s,\e$ correlators, computing bounds on the operator dimensions $(\De_\s,\De_\e)$ and external OPE coefficients $\vec \l = (\l_B,\l_F,\l_{\s\s\e},\l_{\e\e\e},\l_{TT\e})$. We make the (very conservative) gap assumptions listed in table~\ref{tab:gaps}.

\begin{table}
\centering
\begin{tabular}{c|c|c|c|c}
$\Z_2$ & parity & $\ell$ & gap assumption & 3d Ising
\\
\hline
$+$ & $+$ & 0 & $\De\geq 3$ & $\De = 3.82951(61)$ \cite{Reehorst:2021hmp} \\ \hline 
$-$ & $+$ & 0 & $\De \geq 3$ & $\De = 5.262(89)$ \cite{Reehorst:2021hmp} \\ \hline 
$+$ & $-$ & 0 & $\De\geq 3$ & $\De \gtrsim 10$ \cite{Zhu:2022gjc} \\ \hline
$+$ & $+$ & 2 & $\De\geq 4$ & $\De = 5.499(17)$ \cite{Reehorst:2021hmp} \\ \hline 
\multicolumn{2}{l}{otherwise} & & $\tau \geq 1+10^{-6}$ & $\tau \gtrsim 1.0226$ \cite{Simmons-Duffin:2016wlq}
\end{tabular}
\caption{\label{tab:gaps}Our gap assumptions for operators other than $\s,\e,T$, used for computing the allowed island in the 3d Ising CFT\@. Essentially, we assume that all scalars (other than $\s$, $\e$) are irrelevant, and that there is a gap of $1$ above the stress tensor. We also assume a small twist gap in all other sectors ($\tau=\De-\ell$). In the right-hand column, we display the expected actual values for these gaps in the 3d Ising CFT\@. In the last row of the right-hand column, the twist gap comes from the $\Z_2$-even, parity-even, spin-4 sector, which has $\De_{C_{\mu\nu\rho\sigma}}=5.022665(28)$. Clearly our gap assumptions are very conservative.}
\end{table}

At lower derivative orders we computed bounds following the navigator and skydiving approach described above. We found the minimum of the navigator $\mathcal{N}_{\min}$ and maximum and minimum values of $\De_\s$ across the space of external data. For example, at $\Lambda = 11, 19$ we obtained the following results for the $\vec{x}$ vector defined in $(\ref{eq:x})$:

\noindent $\Lambda = 11$:
\begin{align*}
\mathcal N_{\min}: (0.5181484162, 1.412527721, 1.063394763, 0.01494714711, 1.103578160, 1.607303854), \\
\max \De_\s: (0.5182851826, 1.413917985, 1.061046759, 0.01527714250, 1.100931019, 1.606494761),\\
\min \De_\s: (0.5181429340, 1.412553256, 1.063390426, 0.01490640017, 1.103509587, 1.607512121).
\end{align*}
$\Lambda = 19$:
\begin{align*}
\mathcal N_{\min}:  (0.5181487291, 1.412622963, 1.063275748, 0.01491576190, 1.103369155, 1.607484793), \\
\max \De_\s: (0.5181534635, 1.412679013, 1.063183417, 0.01492340043, 1.103256933, 1.607464832), \\
\min \De_\s:  (0.5181429340, 1.412553256, 1.063390426, 0.01490640017, 1.103509587, 1.607512121).
\end{align*}

To compute bounds at higher derivative order, we followed a similar procedure to the computations of bootstrap islands in the $O(2)$ model \cite{Chester:2019ifh}, $O(3)$ model \cite{Chester:2020iyt}, and GNY models \cite{Erramilli:2022kgp}. For each point in $(\De_\s,\De_\e)$-space, we use the OPE ``cutting surface" algorithm of \cite{Chester:2019ifh} to find an allowed direction for the vector $\vec \l$. The output of this algorithm is either an equivalence class $[\vec \l]\in \mathbb{RP}^4$ for which the corresponding SDP is primal feasible, or evidence that no such $[\vec \l]$ exists (see section~\ref{sec:certifying} for details). In the former case, we say that the point $(\De_\s,\De_\e)$ is ``allowed," and in the latter case we say it is ``disallowed."

\begin{figure}
\centering
\includegraphics[width=0.8\textwidth]{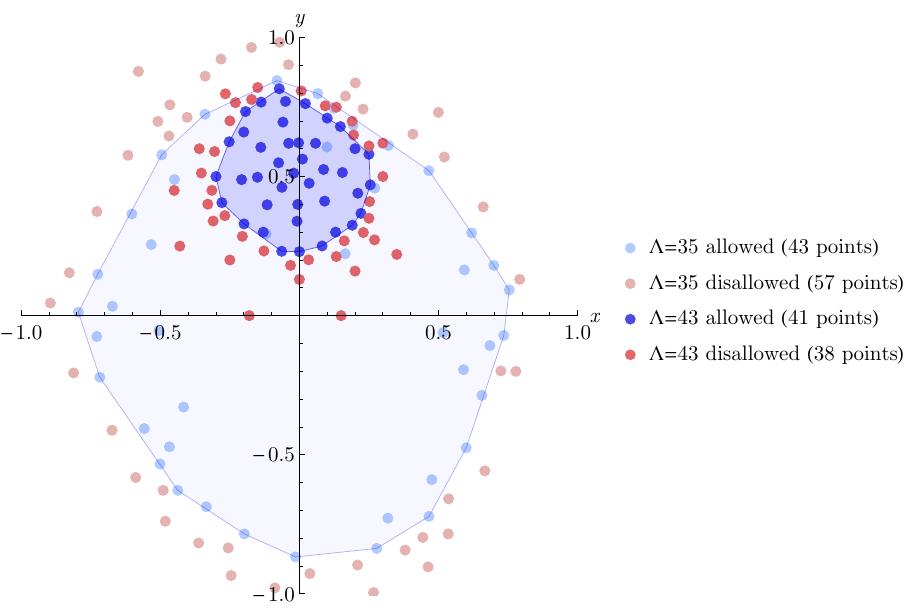}
\caption{\label{fig:affineframe}Allowed and disallowed points with gaps given in table~\ref{tab:gaps}, in the affine frame (\ref{eq:affinecoords}) that makes the allowed islands roughly disc-shaped, computed at $\Lambda=35,43$. Note that disallowed points at $\Lambda=35$ are also disallowed at $\Lambda=43$. Similarly, all allowed points at $\Lambda=43$ are allowed at $\Lambda=35$. To aid visualization, we also display the convex hulls of the allowed points, but note that this is slightly smaller than the true ``allowed region.''}
\end{figure}

We use the Delaunay triangulation search algorithm of \cite{Chester:2019ifh} to choose points in $(\De_\s,\De_\e)$-space to test for allowed/disallowed status, until a reasonably clear picture of the allowed island emerges. To assist the Delaunay search, it is helpful to use affine coordinates where the allowed island is expected to be roughly disc-shaped. We determine such affine coordinates by working our way up in derivative order $\Lambda$.  At a given value of $\Lambda$, we find the allowed region and use principal component analysis to find coordinates that make it roughly disc-shaped. We then use these coordinates for the Delaunay search at the next higher value $\Lambda'>\Lambda$.  As an example, the affine coordinates $(x,y)$ determined from the $\Lambda=35$ allowed region (and used in the $\Lambda=43$ Delaunay search) are defined by
\be
\label{eq:affinecoords}
\De_\s &= 0.5181488106917789+8.496193518377354\cdot 10^{-8} x-5.774118868365222 \cdot 10^{-9} y,
\nn\\
\De_\e &= 1.412625302544141+1.061340020903086 \cdot  10^{-6} x+4.622272819035116 \cdot 10^{-10} y.
\ee
In addition, the Delaunay search requires some initial allowed and disallowed points. We determined these by a combination of guesswork, using results at lower $\Lambda$, and in some cases, results from using skydiving to minimize the GFF navigator function (see section~\ref{sec:navigator}).\footnote{As discussed in section~\ref{sec:navigator}, we chose not to use skydiving to determine the allowed islands at $\Lambda=35$ and $\Lambda=43$ for performance reasons. Skydiving requires building many SDPs corresponding to different values of $(\Delta_\sigma, \Delta_\epsilon)$. Furthermore, these cannot be built in parallel since the algorithm is sequential. Given the cost of building a single SDP, we found it was more efficient to use the methods of \cite{Chester:2019ifh} for our problem.}
Our allowed and disallowed points in $(x,y)$-space for $\Lambda=35,43$ are illustrated in the affine coordinates (\ref{eq:affinecoords}) in figure~\ref{fig:affineframe}.

The allowed islands for $\Lambda=19,27,35,43$ are illustrated in figure~\ref{fig:islands} in the introduction, where we also compare to the allowed island from the $\s$-$\e$ system at $\Lambda=43$~\cite{Kos:2016ysd}.

\begin{figure}
\centering
\includegraphics[width=0.5\textwidth]{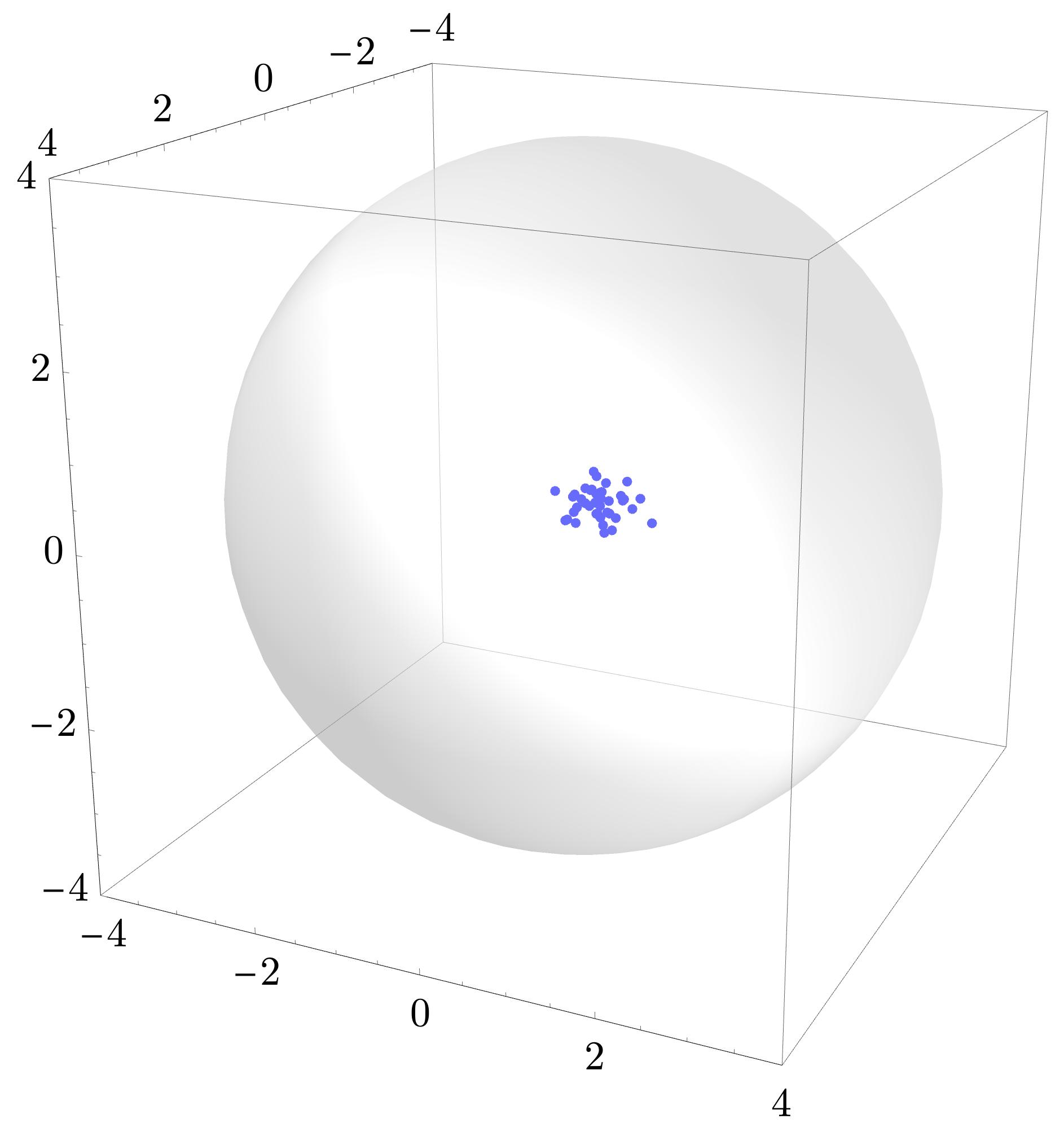}
\caption{\label{fig:boundingellipse}Allowed values of $[\vec \l]$ found at $\Lambda=43$, compared to the initial bounding ellipse, in coordinates where the bounding ellipse is a sphere of radius $4$. Here, we show just 3 out of 4 coordinates of $[\vec \l]$ --- the other coordinate is similar. All the allowed points are deep inside the bounding ellipse.}
\end{figure}

The OPE cutting surface algorithm requires an initial bounding ellipse to work efficiently. We determine it by working our way up in derivative order $\Lambda$, with a method similar to our strategy for finding affine coordinates $(x,y)$. At a given value of $\Lambda$, we collect together all the allowed points in OPE space $[\vec \lambda]$ and use principal component analysis (in an affine patch of $\mathbb{RP}^4$) to find an ellipse that contains them all. Then we expand this ellipse by an $O(1)$ factor, say $4$. Finally, we use the expanded ellipse as the bounding ellipse for the next higher value $\Lambda'>\Lambda$. As a consistency check, we observe afterwards that the allowed OPE points $[\vec \lambda]$ at the higher $\Lambda'$ cluster in a small region deep in the interior of the bounding ellipse, indicating that the results are independent of the choice of bounding ellipse. As an example, the bounding ellipse at $\Lambda=43$ is defined by $\vec \l^T Q_0 \vec \l\leq 0$, where $Q_0$ is given by the matrix
\be
{\scriptsize
\left(
\begin{array}{ccccc}
 0.37440036039553 & 0.37933721733773 & -0.27957642579931 & -0.0020965065873632 & -0.091902746129379 \\
 0.37933721733773 & 0.40707588192755 & -0.28259224067795 & 0.0011378561741937 & -0.099437358761553 \\
 -0.27957642579931 & -0.28259224067795 & 0.20879627066583 & 0.0016234621598708 & 0.068492801656868 \\
 -0.0020965065873632 & 0.0011378561741937 & 0.0016234621598708 & 0.00070008679295711 & -0.00070446545208272 \\
 -0.091902746129379 & -0.099437358761553 & 0.068492801656868 & -0.00070446545208272 & 0.024760700718227 \\
\end{array}
\right).
}
\ee
In figure~\ref{fig:boundingellipse}, we show a plot of the allowed points $[\vec \lambda]$ at $\Lambda=43$, compared to this bounding ellipse (which is a sphere of radius $4$ in the coordinates of the plot). As claimed, all the allowed points are deep in the interior.

As mentioned above, each allowed point in $(\De_\s,\De_\e)$-space comes with an equivalence class $[\vec \lambda]\in \mathbb{RP}^4$, determined by the OPE cutting surface algorithm. Next, for each allowed point, we compute upper and lower bounds on the norm-squared $|\vec \lambda|^2$ using the standard procedure for bounding OPE coefficients \cite{Caracciolo:2009bx}. This results in pairs of OPE coefficient vectors $\vec \l_u, \vec \l_l$ with norms given by the upper/lower bounds on $|\vec \lambda|$. Collecting together the data $(\De_\s,\De_\e,\vec \l_u)$ and $(\De_\s,\De_\e,\vec \l_l)$, we obtain a cloud of allowed points in $\R^7$. (We actually have allowed line segments between each pair of $\vec \l_l$ and $\vec \l_u$, but in practice, $|\vec \l_l-\vec \l_u|$ is  tiny compared to the differences between $\vec \l$ at different points in dimension space, so the distinction is not important.) The bounds on $|\vec \l|^2$ come with extremal functionals containing interesting information about the spectrum of the 3d Ising CFT, which we will report on in future work.

The shape of the cloud of allowed points in $\R^7$ suggests that we can obtain reasonable bounds on the components $(\De_\s,\De_\e,\vec \l)$ by simply taking their minimum and maximum values over the cloud of allowed points (equivalently, using the smallest box in $\R^7$ that is aligned with the axes and contains all the allowed points).\footnote{To make sure that our cloud of allowed points is representative, we sampled a number of points deep in the interior of the $\L=43$ allowed region in the dimension space, see figure~\ref{fig:affineframe}.} This leads to the bounds reported in table~\ref{tab:mainresults}. Let us explain this in more detail. First, note that our cloud of allowed points in $\R^7$ is long and thin: the eigenvalues of its covariance matrix have relative sizes
\be
(1,0.0493,0.0144,0.00261,0.00165,0.000555,0.000126),
\ee
where we have divided by the largest eigenvalue. This suggests that the allowed region in $\R^7$ can be well-approximated by a low-dimensional shape, say a 2-dimensional shape. Furthermore, by the Delaunay search, the known allowed points approximately cover the projection of this 2-dimensional shape to $(\De_\s,\De_\e)$-space. This suggests that the known allowed points in $\R^7$ give a good approximation of the allowed region. The relative error can be estimated using the 3rd and higher eigenvalues of the covariance matrix, which is $\lesssim 0.3 \%$ compared to the size of the island.

\subsubsection{The parity-odd sector of the Ising CFT}

\begin{figure}
\centering
\includegraphics[width=0.89\textwidth]{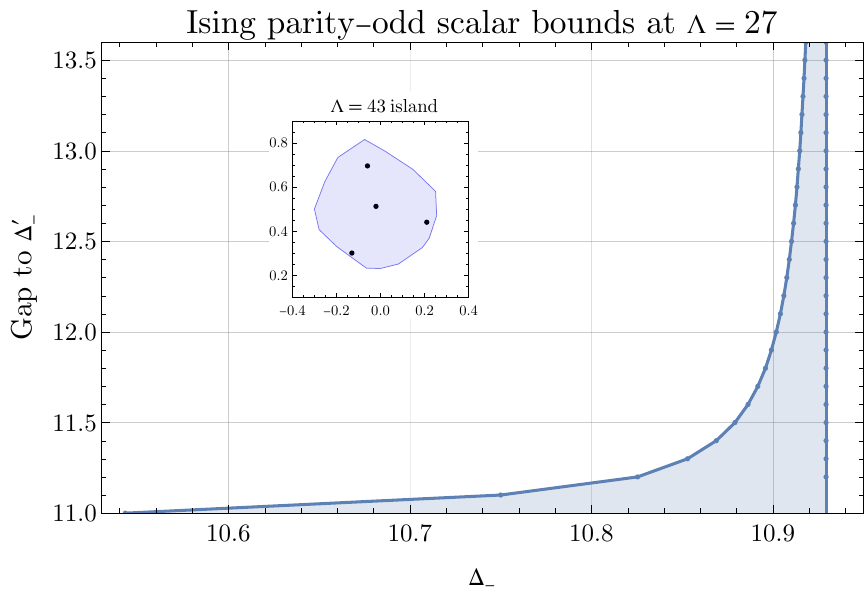}
\caption{\label{fig:delta_odd_prime} Bounds on scaling dimensions of the two lightest parity-even, $\Z_2$-odd scalars in the Ising model, computed at $\Lambda=27$. The lower bound on $\De_-$ is sensitive to the value of $\De_-'$, while the upper bound is not. The region shown here is the union of allowed regions for several representative points in the $\Lambda=43$ $(\De_\s, \De_\e)$ island, as highlighted on the plot inset.}
\end{figure}

Because $T\times T$ contains parity-odd tensor structures, the $T,\s,\e$ mixed correlator system can be used to probe parity-odd scalars, giving us access to a new sector of operators. The $\Z_2$-even, parity-odd scalar channel has an unusually large gap. For intuition, note that the lightest analogous operator in the free scalar CFT is
\be
\mathcal O_- = \; :\! \e^{\mu\nu\rho} \phi (\partial_\alpha \partial_{\beta_1} \partial_{\beta_2} \partial_\mu \phi) (\partial^\alpha \partial_\nu \phi) (\partial^{\beta_1} \partial^{\beta_2} \partial_\rho \phi)\!:,
\ee
which has scaling dimension $\De_- = 11$. The second-lightest operator in the free scalar theory, which contains an additional factor of $\phi^2$, has scaling dimension $\De_{-}' = 12$.

We obtain an upper bound on $\De_-$ in the 3d Ising CFT by starting from the spectrum assumptions in table~\ref{tab:gaps} and maximizing the  $\Z_2$-even, parity-odd scalar gap via a binary search. At each value of the gap, we perform an OPE scan over the bounding ellipse depicted in 3d. Because the maximum gap depends on the operator dimensions $(\De_\s, \De_\e)$, we repeat this process for a few points in the final $\Lambda=43$  Ising island (figure~\ref{fig:affineframe}; our search points are indicated on the inset of figure ~\ref{fig:delta_odd_prime}). This allows us to incorporate all available information about the dimensions while running the search at a much lower derivative order. At $\Lambda=27$, we find an upper bound of
\be
\De_- \le 10.9293
\ee
which very nearly coincides with the universal bound of $\Delta_-\lesssim 10.97$ (figure~\ref{fig:tttt_scalars_ising}).  This lends support to the conjecture that the 3d Ising model saturates this universal bound.

It's worth noting that all trial values of  $(\De_\s, \De_\e)$ produce identical upper bounds, at least within the binary search stopping tolerance of $10^{-4}$. This is expected, as the $\Lambda=43$ island covers roughly 1\% of the area of the  $\Lambda=27$ one, so all points in it should look roughly the same from the perspective of the $\Lambda=27$ crossing equations.

To compute a \emph{lower} bound on $\De_-$, we must impose a gap above this operator, which requires us to make some (physically motivated) assumptions about $\De_-'$.  It is reasonable to guess that $\De_-'$ is close to the free theory value of 12. A conservative gap of $\De_-' \ge 11$ implies a lower bound of $\De_- \ge 10.5432$, and a less conservative gap of $\De_-' \ge 11.5$ implies a lower bound of $\De_- \ge 10.8790$.
An exclusion plot in $(\De_-,\De_-')$-space is illustrated in figure~\ref{fig:delta_odd_prime}. As in the gap maximization searches, all points in $(\De_\s, \De_\e)$-space return similar lower bounds---the largest discrepancy between bounds from different points is 0.0002.

These lower bounds are considerably higher than fuzzy sphere estimate of $\De_- = 10.01(30)$ obtained in \cite{Zhu:2022gjc} (Note that the authors do not quantify the uncertainty of this operator's scaling dimension, but suggest that a 3\% relative uncertainty is reasonable based on the uncertainties of other primaries' scaling dimensions). By fixing $\De_-$ to the fuzzy sphere value and scanning over various values of the $\De_-'$ gap, we find that it requires $\De_-' \le 10.95$.

\section{Conclusions}
\label{sec:conclusions}

Our results give very precise determinations of leading scaling dimensions and OPE coefficients of the 3d Ising CFT, including data involving the stress tensor. We expect that one can learn about other operators in the spectrum by extracting and analyzing the extremal functionals \cite{Poland:2010wg,El-Showk:2012vjm} corresponding to various boundaries of the 3d Ising island. This data should include trajectories of double-twist operators built out of the stress tensor which have so far not been studied in the numerical bootstrap. It will be very interesting to analyze these trajectories, comparing with predictions from the Lorentzian inversion formula \cite{Caron-Huot:2017vep}, in the spirit of \cite{Simmons-Duffin:2016wlq,Caron-Huot:2020ouj,Liu:2020tpf,Atanasov:2022bpi}. This data will also give access to interesting observables associated with the ANEC operator in the 3d Ising CFT, such as energy correlators via the light-ray OPE \cite{Hofman:2008ar,Kologlu:2019mfz}.

One amusing interpretation of our determinations of \(\<TTT\>\) OPE coefficients is that the 3d Ising CFT is ``\(n_F /(n_F+n_B)=1.3833683(35)\%\) fermionic". We may wonder how this percentage compares against other interacting CFTs, such as the critical \(O(N)\) vector models or CFTs built from interacting fermions or gauge fields. In particular, how these OPE coefficients may vary across different families of CFTs could well be nontrivial in light of conjectured bosonization/fermionization dualities \cite{Seiberg:2016gmd}; see \cite{Turner:2019wnh} for an excellent pedagogical review. Bootstrapping other mixed systems involving the stress tensor therefore would be an interesting line of inquiry. We leave these exciting directions to future work.

In section~\ref{sec:ttttresults}, we found that even the $\<TTTT\>$ system alone can produce sharp features corresponding to the 3d Ising CFT\@. This observation leads to further questions: are there more such features in $\<TTTT\>$ bounds, and can they be related to other physical CFTs? (Note that the stress tensor bootstrap automatically excludes non-local CFTs, such as the long-range Ising model \cite{Behan:2017dwr,Behan:2017emf}.)
Concretely, there is a large region of parameter space to the interior of the universal bounds in figure \ref{fig:tttt_scalars} in which any local, unitary CFT in 3d must exist. Is it possible to map out this interior (and the analogous space for global symmetry currents) by extremizing other parameters? Preliminary explorations reveal that there is a rich structure here and we plan to continue studying it in future work.

Given the effectiveness of including $T_{\mu\nu}$ for increasing the precision of bounds on the 3d Ising CFT, it is natural to ask: which operator is next? That is, which additional operator(s) should we include in the crossing equations to further isolate this theory? We do not have a rigorous answer to this question, but intuition from the lightcone bootstrap \cite{Simmons-Duffin:2016wlq,Su:2022xnj,Li:2023tic} suggests that low-twist operators could play an important role. Note that $\sigma$ is the lowest-twist operator in the 3d Ising CFT, and $T_{\mu\nu}$ is the next-lowest. After that, we have the leading spin-4 operator $C_{\mu\nu\rho\s}$, with twist approximately $\tau_C\approx 1.022$. However, it is daunting (from the point of view of numerics) to imagine the system of crossing equations we would obtain by studying four-point functions of $C_{\mu\nu\rho\s}$ and other operators. As an example, the number of parity-even four-point structures in the $\<CCCC\>$ four-point function alone is 881, which can be contrasted with the 29 ``bulk" crossing equations studied in this work. It is interesting to ask whether we can include only the low-twist components of $C_{\mu\nu\rho\s}$ in an efficient way. An alternative approach is to consider higher-point crossing equations, which encode an infinite number of four-point crossing equations at once. See \cite{Poland:2023vpn,Poland:2023bny,Harris:2024nmr,Antunes:2023kyz} for recent work in this direction.

\section*{Acknowledgements}
We thank  Yin-Chen He, Johan Henriksson, Christopher Herzog, Simon Caron-Huot, Luca Iliesiu, Denis Karateev, Walter Landry, Valentina Prilepina, Leonardo Rastelli, Slava Rychkov, Andreas Stergiou, Ning Su, Petar Tadi\'{c}, Alessandro Vichi, and Xi Yin for discussions. 

DP and MM were supported by Simons Foundation grant 488651 (Simons Collaboration on the Nonperturbative Bootstrap) and DOE grant DE-SC0017660. AH is supported by Simons Foundation grant 994312 (Simons Collaboration on Confinement and QCD Strings). The work of PK was funded by UK Research and Innovation (UKRI) under the UK government's Horizon Europe funding Guarantee [grant number EP/X042618/1] and the Science and Technology Facilities Council [grant number  ST/X000753/1]. Research at the Perimeter Institute is supported in part by the Government of Canada through NSERC and by the Province of Ontario through MRI. This research was supported in part by grant NSF PHY-2309135 to the Kavli Institute for Theoretical Physics (KITP).  This material is based upon work supported by the U.S.\
Department of Energy, Office of Science, Office of High Energy Physics, under Award Number DE-SC0011632. CHC, VD, AL, and DSD are supported in part by Simons Foundation grant 488657 (Simons Collaboration on the Nonperturbative Bootstrap). RSE is supported by Simons Foundation grant 915279 (IHES).

This work used the following computational facilities:
\begin{itemize}
	\item Expanse cluster at the San Diego Supercomputing Center (SDSC) through allocation PHY190023 from the Advanced Cyberinfrastructure Coordination Ecosystem: Services \& Support (ACCESS) program, which is supported by National Science Foundation grants \#2138259, \#2138286, \#2138307, \#2137603, and \#2138296. 
	\item DiRAC Memory Intensive service Cosma8 at Durham University, managed by the Institute for Computational Cosmology on behalf of the STFC DiRAC HPC Facility (www.dirac.ac.uk). The DiRAC service at Durham was funded by BEIS, UKRI and STFC capital funding, Durham University and STFC operations grants. DiRAC is part of the UKRI Digital Research Infrastructure.
	\item Resnick High Performance Computing Center, a facility supported by Resnick Sustainability Institute at the California Institute of Technology.
	\item Yale Grace computing cluster, supported by the facilities and staff of the Yale University Faculty of Sciences High Performance Computing Center.
	\item CREATE High Performance Cluster~\cite{CREATE} at King's College London.
\end{itemize}

\appendix

\section{Three-point structures}
\label{app:threept}

In this appendix, we record our choice of three-point structures for correlators of $\s,\e,T$ in the 3d Ising CFT\@. We follow the conventions of \cite{Erramilli:2022kgp}, where we provide a three-point structure for each ordered triple of operators that can appear in a three-point correlator $\<\cO_i\cO_j\cO_k\>$. These structures must be chosen so that the OPE coefficients $\l^a_{ijk}$ are functions of the {\it unordered\/} triple $(i,j,k)$.\footnote{In practice this means that the structures are related by permutations. By providing the functions explicitly, we allow the code that prepares conformal block expansions to be independent of the details of how permutations act on local operators.} When all three operators are external (i.e.\ $\s,\e,$ or $T$), we specify structures for all unique ordered triples. When one of the operators is internal, we choose conventions where the internal operator appears last, so we only specify structures for up to two unique orderings.

In the tables below, each section bounded by horizontal lines corresponds to an independent OPE coefficient $\l_{ijk}$. Within each section, each row corresponds to an ordered triple of operators and its corresponding three-point structure. The notation $|j_{12},j_{123}\>$ denotes a conformal structure in the $\SO(3)_r$ basis of \cite{Erramilli:2019njx,Erramilli:2020rlr}. In addition, we also have specialized three-point structures such as $\texttt{StressTensorB}$, $\texttt{ScalarParityEven}$, etc., which are defined in the ancillary file \texttt{three\_point\_structures.m}.

In the tables below, we work in conventions where the two-point function of stress tensors is unit-normalized. Thus, $T$ in this section is the same as $\hat T$ in section~\ref{sec:ward}.

\paragraph{External operators}

There are five OPE coefficients corresponding to external operators: $(\l_B,\l_F,\l_{\s\s\e},\l_{\e\e\e},\l_{TT\e})$. Due to the Ward identity (\ref{eq:nicewardforscalars}), the OPE coefficients $\l_B$ and $\l_F$ appear in three-point structures for triples of multiple different operators.

\begin{align}
\label{eq:externalopes}
\begin{array}{|l|l|l|}
\hline
\textrm{OPE coefficient} & \textrm{ordered triple $\<\cO_i\cO_j\cO_j\>$} & \textrm{three-point structure}\\
\hline
\l_B &\<T T T\> & \texttt{StressTensorB}\\
& \<T\sigma\sigma\> &  - \Delta_\sigma |2,2\rangle\\
&\<\sigma T \sigma\> &  - \Delta_\sigma |2,2\rangle\\
&\<\sigma \sigma T\> &  - \Delta_\sigma |0,2\rangle\\
&\<T\epsilon\epsilon\> &  - \Delta_\epsilon  |2,2\rangle\\
&\<\epsilon T \epsilon\> &  - \Delta_\epsilon |2,2\rangle\\
&\<\epsilon \epsilon T\> &  - \Delta_\epsilon |0,2\rangle\\
\hline
\l_F &\<T T T\> & \texttt{StressTensorF}\\
&\<T \sigma \sigma\> &  - \Delta_\sigma |2,2\rangle\\
&\<\sigma T \sigma\> &  - \Delta_\sigma  |2,2\rangle\\
&\<\sigma \sigma T\> &  - \Delta_\sigma |0,2\rangle\\
&\<T \epsilon \epsilon\> &  - \Delta_\epsilon  |2,2\rangle\\
&\<\epsilon T \epsilon\> &  - \Delta_\epsilon  |2,2\rangle\\
&\<\epsilon \epsilon T\> &  - \Delta_\epsilon  |0,2\rangle\\
\hline
\l_{\s\s\e} &\<\sigma \sigma \epsilon\> & |0,0\rangle\\
&\<\sigma \epsilon \sigma\> & |0,0\rangle\\
&\<\epsilon \sigma \sigma\> & |0,0\rangle\\
\hline
\l_{\e\e\e} &\<\epsilon \epsilon \epsilon\> & |0,0\rangle\\
\hline
\l_{TT\e} &\<T T \epsilon\> &  -\frac{16\De_\e}{2 + \De_\e}\ \texttt{ScalarParityEven}\\
&\<T \epsilon T\> & \frac{6}{\De_\e(2 + \De_\e)}\ \texttt{TSOParityEven}_2\\
&\<\epsilon T T\> & \frac{6}{\De_\e(2 + \De_\e)}\ \texttt{STOParityEven}_2\\
 \hline\end{array}\end{align}

In subsequent tables, we omit the OPE coefficient names (i.e.\ the left-most column in the table above).

\paragraph{$\Z_2$-even, parity-even, $j = 0$}
\begin{align}\begin{array}{|l|l|}\hline\<T T \cO\> & \texttt{ScalarParityEven}\\
\hline
\<\sigma \sigma \cO\> & |0,0\rangle\\
\hline
\<\epsilon \epsilon \cO\> & |0,0\rangle\\ \hline\end{array}\end{align}

\paragraph{$\Z_2$-even, parity-even, $j = 2$}
\begin{align}\begin{array}{|l|l|}\hline\<T T\cO\> & \texttt{Spin2ParityEven}\\
\hline
\<\sigma \sigma \cO\> & |0,2\rangle\\
\hline
\<\epsilon \epsilon \cO\> & |0,2\rangle\\
\hline
\<T \epsilon \cO\> & \texttt{TSOParityEven}_2\\
\<\epsilon T \cO\> & \texttt{STOParityEven}_2\\ \hline\end{array}\end{align}

\paragraph{$\Z_2$-even, parity-even, $j\geq 3$ odd}
\begin{align}\begin{array}{|l|l|}\hline\<T \epsilon \cO\> & \texttt{TSOParityEven}_j\\
\<\epsilon T \cO\> & \texttt{STOParityEven}_j\\ \hline\end{array}\end{align}

\paragraph{$\Z_2$-even, parity-even, $j \geq 4$ even}
\begin{align}\begin{array}{|l|l|}\hline
\<T T \cO\> & \texttt{GenericParityEven1}_j\\
\hline
\<T T \cO\> & \texttt{GenericParityEven2}_j\\
\hline
\<\sigma \sigma \cO\> & |0,j\rangle\\
\hline
\<\epsilon \epsilon \cO\> & |0,j\rangle\\
\hline
\<T \epsilon \cO\> & \texttt{TSOParityEven}_j\\
\<\epsilon T \cO\> & \texttt{STOParityEven}_j\\ \hline\end{array}\end{align}

\paragraph{$\Z_2$-even, parity-odd, $j = 0$}
\begin{align}\begin{array}{|l|l|}\hline\<T T \cO\> & \texttt{ScalarParityOdd}\\ \hline\end{array}\end{align}

\paragraph{$\Z_2$-even, parity-odd, $j = 2$}
\begin{align}\begin{array}{|l|l|}\hline\<T T \cO\> & \texttt{Spin2ParityOdd}\\
\hline
\<T \epsilon \cO\> & \texttt{TSOParityOdd}_2\\
\<\epsilon T \cO\> & \texttt{STOParityOdd}_2\\ \hline\end{array}\end{align}

\paragraph{$\Z_2$-even, parity-odd, $j = 3$}
\begin{align}\begin{array}{|l|l|}\hline\<T \epsilon \cO\> & \texttt{TSOParityOdd}_3\\
\<\epsilon T \cO\> & \texttt{STOParityOdd}_3\\ \hline\end{array}\end{align}

\paragraph{$\Z_2$-even, parity-odd, $j \geq 4$}
\begin{align}\begin{array}{|l|l|}\hline\<T T \cO\> & \texttt{GenericParityOdd}_j\\
\hline
\<T \epsilon \cO\> & \texttt{TSOParityOdd}_j\\
\<\epsilon T \cO\> & \texttt{STOParityOdd}_j\\ \hline\end{array}\end{align}

\paragraph{$\Z_2$-odd, parity-even, $j = 0$}
\begin{align}\begin{array}{|l|l|}\hline\<\sigma \epsilon \cO\> & |0,0\rangle\\
\<\epsilon \sigma \cO\> & |0,0\rangle\\ \hline\end{array}\end{align}

\paragraph{$\Z_2$-odd, parity-even, $j = 1$}
\begin{align}\begin{array}{|l|l|}\hline\<\sigma \epsilon \cO\> & |0,1\rangle\\
\<\epsilon \sigma \cO\> &  - |0,1\rangle\\ \hline\end{array}\end{align}

\paragraph{$\Z_2$-odd, parity-even, $j \geq 2$}
\begin{align}\begin{array}{|l|l|}\hline\<T \sigma \cO\> & \texttt{TSOParityEven}_j\\
\<\sigma T \cO\> & \texttt{STOParityEven}_j\\
\hline
\<\sigma \epsilon \cO\> & |0,j\rangle\\
\<\epsilon \sigma \cO\> & (-1)^j|0,j\rangle\\ \hline\end{array}\end{align}

\paragraph{$\Z_2$-odd, parity-odd, $j \geq 2$}
\begin{align}\begin{array}{|l|l|}\hline\<T \sigma \cO\> & \texttt{TSOParityOdd}_j\\
\<\sigma T \cO\> & \texttt{STOParityOdd}_j\\ \hline\end{array}\end{align}

\section{Crossing equations for the $T\sigma\epsilon$ system}
\label{sec:appendix_crossing}

\newcommand\nmax{n_\textrm{max}}

In this appendix, we present the crossing equations for correlators of $T$, $\s$, and $\e$. We use the cross-ratios $w,s$ defined by  \cite{Mazac:2016qev,Erramilli:2020rlr}
\be
z = \frac{(1+y)^2}{2(1+y^2)},\quad \bar z = \frac{(1+\bar y)^2}{2(1+\bar y^2)}, \nn\\
w = \frac{y+\bar y}{2},\quad s = \p{\frac{y-\bar y}{2}}^2.
\ee
In these coordinates, the crossing symmetric point is $w=s=0$. We denote evaluation of a function $g(w,s)$ at the crossing symmetric point by $g|_* \equiv g(w=0,s=0) $. The careful reader may worry that with this definition of $s$ we omit structures that are odd under \(z\leftrightarrow\bar z\); as noted in section \ref{subsec:4pt}, only the even structures appear. For more general cases where this may not be true, refer to \cite{Erramilli:2022kgp}.

\begin{table}
\centering
\begin{tabular}{c|c|c}
$\Lambda$ & $\nmax$ & components \\
\hline
11 & 6 & 743\\
\hline
19 & 10 & 1835\\
\hline
27 & 14 & 3407\\
\hline
35 & 18 & 5459\\
\hline
43 & 22 & 7991
\end{tabular}
\caption{\label{tab:numcomponents}Number of components of the functional acting on crossing equations for the $T\s\e$ system, for different derivative orders.}
\end{table}

The number of derivatives $\ptl_w^m \ptl_s^n$ of each crossing equation is determined by an integer parameter $\Lambda$. It is also useful to define $\nmax=\lfloor{\frac{\Lambda+1}{2}}\rfloor$. We use the following terminology:
\begin{itemize}
\item ``bulk equations" are equations that involve derivatives in both cross-ratios $w$ and $s$. In our conventions, we symmetrize/anti-symmetrize each equation with respect to the crossing transformation $w\to -w$. Thus, each bulk equation involves either even or odd derivatives with respect to $w$. Consequently, each bulk equation has $\frac{\nmax(\nmax+1)}{2}$ components.
\item ``line equations" (or ``radial equations") are equations that involve derivatives only along the line $s=0$. Each line equation has $\nmax$ components.
\item ``point equations" involve no derivatives, and are given by evaluating an expression at the crossing-symmetric point. Each point equation has $1$ component.
\end{itemize}
Overall, the $T\s\e$ system has 30 bulk equations, 18 line equations, and 5 point equations, for a total of
\be
\dim \vec \a &= 30 \frac{\nmax(\nmax+1)}{2} + 18 \nmax + 5
\ee
functional components, see table~\ref{tab:numcomponents}. The crossing equations are listed below. In each equation $m,n\geq 0$ are nonnegative integers subject to the constraint $m+2n\leq \Lambda$.

\paragraph{Bulk equations for $\<TTTT\>$}
\begin{align}
0 &= \left.\partial_w^m \partial_s^n \left(g_{T T T T}^{\<2,2,2,2\>^+}\right)\right|_{*}, && (\textrm{$m$ odd})\nonumber\\
0 &= \left.\partial_w^m \partial_s^n \left(g_{T T T T}^{\<1,1,1,1\>^+}\right)\right|_{*}, && (\textrm{$m$ odd})\nonumber\\
0 &= \left.\partial_w^m \partial_s^n \left(g_{T T T T}^{\<1,1,2,2\>^+} + g_{T T T T}^{\<2,1,1,2\>^+}\right)\right|_{*}, && (\textrm{$m$ odd})\nonumber\\
0 &= \left.\partial_w^m \partial_s^n \left(g_{T T T T}^{\<1,1,2,2\>^+} - g_{T T T T}^{\<2,1,1,2\>^+}\right)\right|_{*}, && (\textrm{$m$ even})\nonumber\\
0 &= \left.\partial_w^m \partial_s^n \left(g_{T T T T}^{\<1,2,1,2\>^+}\right)\right|_{*}. && (\textrm{$m$ odd})
\end{align}
\paragraph{Line equations for $\<TTTT\>$}
\begin{align}
0 &= \left.\partial_w^m \left(g_{T T T T}^{\<0,0,0,0\>^+}\right)\right|_{*}, && (\textrm{$m$ odd})\nonumber\\
0 &= \left.\partial_w^m \left(g_{T T T T}^{\<0,1,0,1\>^+}\right)\right|_{*}, && (\textrm{$m$ odd})\nonumber\\
0 &= \left.\partial_w^m \left(g_{T T T T}^{\<0,2,0,2\>^+}\right)\right|_{*}, && (\textrm{$m$ odd})\nonumber\\
0 &= \left.\partial_w^m \left(g_{T T T T}^{\<0,1,1,2\>^+} + g_{T T T T}^{\<1,1,0,2\>^+}\right)\right|_{*}, && (\textrm{$m$ odd})\nonumber\\
0 &= \left.\partial_w^m \left(g_{T T T T}^{\<0,1,1,2\>^+} - g_{T T T T}^{\<1,1,0,2\>^+}\right)\right|_{*}, && (\textrm{$m$ even})\nonumber\\
0 &= \left.\partial_w^m \left(g_{T T T T}^{\<0,0,1,1\>^+} + g_{T T T T}^{\<1,0,0,1\>^+}\right)\right|_{*}, && (\textrm{$m$ odd})\nonumber\\
0 &= \left.\partial_w^m \left(g_{T T T T}^{\<0,0,1,1\>^+} - g_{T T T T}^{\<1,0,0,1\>^+}\right)\right|_{*}, && (\textrm{$m$ even})\nonumber\\
0 &= \left.\partial_w^m \left(g_{T T T T}^{\<-1,0,0,1\>^+} + g_{T T T T}^{\<0,0,-1,1\>^+}\right)\right|_{*}, && (\textrm{$m$ odd})\nonumber\\
0 &= \left.\partial_w^m \left(g_{T T T T}^{\<-1,0,0,1\>^+} - g_{T T T T}^{\<0,0,-1,1\>^+}\right)\right|_{*}. && (\textrm{$m$ even})
\end{align}
\paragraph{Point equations for $\<TTTT\>$}
\begin{align}
0 &= \left.\left(g_{T T T T}^{\<0,0,2,2\>^+} - g_{T T T T}^{\<2,0,0,2\>^+}\right)\right|_{*}, && \nonumber\\
0 &= \left.\left(g_{T T T T}^{\<0,-1,1,2\>^+} - g_{T T T T}^{\<1,-1,0,2\>^+}\right)\right|_{*}, && \nonumber\\
0 &= \left.\left(g_{T T T T}^{\<-1,-1,1,1\>^+} - g_{T T T T}^{\<1,-1,-1,1\>^+}\right)\right|_{*}, && \nonumber\\
0 &= \left.\left(g_{T T T T}^{\<-1,1,0,2\>^+} - g_{T T T T}^{\<0,1,-1,2\>^+}\right)\right|_{*}. &&
\end{align}
\paragraph{Bulk equations for $\<TTT\e\>$}
\begin{align}
0 &= \left.\partial_w^m \partial_s^n \left(g_{T T T \epsilon}^{\<2,2,2,0\>^+}\right)\right|_{*}, && (\textrm{$m$ odd})\nonumber\\
0 &= \left.\partial_w^m \partial_s^n \left(g_{T T T \epsilon}^{\<1,2,1,0\>^+}\right)\right|_{*}, && (\textrm{$m$ odd})\nonumber\\
0 &= \left.\partial_w^m \partial_s^n \left(g_{T T T \epsilon}^{\<1,1,2,0\>^+} + g_{T T T \epsilon}^{\<2,1,1,0\>^+}\right)\right|_{*}, && (\textrm{$m$ odd})\nonumber\\
0 &= \left.\partial_w^m \partial_s^n \left(g_{T T T \epsilon}^{\<1,1,2,0\>^+} - g_{T T T \epsilon}^{\<2,1,1,0\>^+}\right)\right|_{*}. && (\textrm{$m$ even})
\end{align}
\paragraph{Line equations for $\<TTT\e\>$}
\begin{align}
0 &= \left.\partial_w^m \left(g_{T T T \epsilon}^{\<0,0,2,0\>^+} + g_{T T T \epsilon}^{\<2,0,0,0\>^+}\right)\right|_{*}, && (\textrm{$m$ odd})\nonumber\\
0 &= \left.\partial_w^m \left(g_{T T T \epsilon}^{\<0,0,2,0\>^+} - g_{T T T \epsilon}^{\<2,0,0,0\>^+}\right)\right|_{*}, && (\textrm{$m$ even})\nonumber\\
0 &= \left.\partial_w^m \left(g_{T T T \epsilon}^{\<0,2,0,0\>^+}\right)\right|_{*}. && (\textrm{$m$ odd})
\end{align}
\paragraph{Point equations for $\<TTT\e\>$}
\begin{align}
0 &= \left.\left(g_{T T T \epsilon}^{\<-2,2,2,0\>^+} - g_{T T T \epsilon}^{\<2,2,-2,0\>^+}\right)\right|_{*}. &&
\end{align}
\paragraph{Bulk equations for $\<TT\s\s\>$}
\begin{align}
0 &= \left.\partial_w^m \partial_s^n \left(g_{T T \sigma \sigma}^{\<2,2,0,0\>^+} + g_{\sigma T T \sigma}^{\<0,2,2,0\>^+}\right)\right|_{*}, && (\textrm{$m$ odd})\nonumber\\
0 &= \left.\partial_w^m \partial_s^n \left(g_{T T \sigma \sigma}^{\<2,2,0,0\>^+} - g_{\sigma T T \sigma}^{\<0,2,2,0\>^+}\right)\right|_{*}, && (\textrm{$m$ even})\nonumber\\
0 &= \left.\partial_w^m \partial_s^n \left(g_{T T \sigma \sigma}^{\<1,1,0,0\>^+} + g_{\sigma T T \sigma}^{\<0,1,1,0\>^+}\right)\right|_{*}, && (\textrm{$m$ odd})\nonumber\\
0 &= \left.\partial_w^m \partial_s^n \left(g_{T T \sigma \sigma}^{\<1,1,0,0\>^+} - g_{\sigma T T \sigma}^{\<0,1,1,0\>^+}\right)\right|_{*}, && (\textrm{$m$ even})\nonumber\\
0 &= \left.\partial_w^m \partial_s^n \left(g_{\sigma T \sigma T}^{\<0,2,0,2\>^+}\right)\right|_{*}, && (\textrm{$m$ odd})\nonumber\\
0 &= \left.\partial_w^m \partial_s^n \left(g_{\sigma T \sigma T}^{\<0,1,0,1\>^+}\right)\right|_{*}. && (\textrm{$m$ odd})
\end{align}
\paragraph{Line equations for $\<TT\s\s\>$}
\begin{align}
0 &= \left.\partial_w^m \left(g_{T T \sigma \sigma}^{\<0,0,0,0\>^+} + g_{\sigma T T \sigma}^{\<0,0,0,0\>^+}\right)\right|_{*}, && (\textrm{$m$ odd})\nonumber\\
0 &= \left.\partial_w^m \left(g_{T T \sigma \sigma}^{\<0,0,0,0\>^+} - g_{\sigma T T \sigma}^{\<0,0,0,0\>^+}\right)\right|_{*}, && (\textrm{$m$ even})\nonumber\\
0 &= \left.\partial_w^m \left(g_{\sigma T \sigma T}^{\<0,0,0,0\>^+}\right)\right|_{*}. && (\textrm{$m$ odd})
\end{align}
\paragraph{Bulk equations for $\<TT\e\e\>$}
\begin{align}
0 &= \left.\partial_w^m \partial_s^n \left(g_{T T \epsilon \epsilon}^{\<2,2,0,0\>^+} + g_{\epsilon T T \epsilon}^{\<0,2,2,0\>^+}\right)\right|_{*}, && (\textrm{$m$ odd})\nonumber\\
0 &= \left.\partial_w^m \partial_s^n \left(g_{T T \epsilon \epsilon}^{\<2,2,0,0\>^+} - g_{\epsilon T T \epsilon}^{\<0,2,2,0\>^+}\right)\right|_{*}, && (\textrm{$m$ even})\nonumber\\
0 &= \left.\partial_w^m \partial_s^n \left(g_{T T \epsilon \epsilon}^{\<1,1,0,0\>^+} + g_{\epsilon T T \epsilon}^{\<0,1,1,0\>^+}\right)\right|_{*}, && (\textrm{$m$ odd})\nonumber\\
0 &= \left.\partial_w^m \partial_s^n \left(g_{T T \epsilon \epsilon}^{\<1,1,0,0\>^+} - g_{\epsilon T T \epsilon}^{\<0,1,1,0\>^+}\right)\right|_{*}, && (\textrm{$m$ even})\nonumber\\
0 &= \left.\partial_w^m \partial_s^n \left(g_{\epsilon T \epsilon T}^{\<0,2,0,2\>^+}\right)\right|_{*}, && (\textrm{$m$ odd})\nonumber\\
0 &= \left.\partial_w^m \partial_s^n \left(g_{\epsilon T \epsilon T}^{\<0,1,0,1\>^+}\right)\right|_{*}. && (\textrm{$m$ odd})
\end{align}
\paragraph{Line equations for $\<TT\e\e\>$}
\begin{align}
0 &= \left.\partial_w^m \left(g_{T T \epsilon \epsilon}^{\<0,0,0,0\>^+} + g_{\epsilon T T \epsilon}^{\<0,0,0,0\>^+}\right)\right|_{*}, && (\textrm{$m$ odd})\nonumber\\
0 &= \left.\partial_w^m \left(g_{T T \epsilon \epsilon}^{\<0,0,0,0\>^+} - g_{\epsilon T T \epsilon}^{\<0,0,0,0\>^+}\right)\right|_{*}, && (\textrm{$m$ even})\nonumber\\
0 &= \left.\partial_w^m \left(g_{\epsilon T \epsilon T}^{\<0,0,0,0\>^+}\right)\right|_{*}. && (\textrm{$m$ odd})
\end{align}
\paragraph{Bulk equations for $\<T\e\e\e\>$}
\begin{align}
0 &= \left.\partial_w^m \partial_s^n \left(g_{\epsilon T \epsilon \epsilon}^{\<0,2,0,0\>^+}\right)\right|_{*}. && (\textrm{$m$ odd})
\end{align}
\paragraph{Bulk equations for $\<\s\s\e T\>$}
\begin{align}
0 &= \left.\partial_w^m \partial_s^n \left(g_{\sigma T \epsilon \sigma}^{\<0,2,0,0\>^+} + g_{\epsilon T \sigma \sigma}^{\<0,2,0,0\>^+}\right)\right|_{*}, && (\textrm{$m$ odd})\nonumber\\
0 &= \left.\partial_w^m \partial_s^n \left(g_{\sigma T \epsilon \sigma}^{\<0,2,0,0\>^+} - g_{\epsilon T \sigma \sigma}^{\<0,2,0,0\>^+}\right)\right|_{*}, && (\textrm{$m$ even})\nonumber\\
0 &= \left.\partial_w^m \partial_s^n \left(g_{\sigma T \sigma \epsilon}^{\<0,2,0,0\>^+}\right)\right|_{*}. && (\textrm{$m$ odd})
\end{align}
\paragraph{Bulk equations for the $\s\e$ system \cite{Kos:2014bka}}
\begin{align}
0 &= \left.\partial_w^m \partial_s^n \left(g_{\sigma \sigma \sigma \sigma}^{\<0,0,0,0\>^+}\right)\right|_{*}, && (\textrm{$m$ odd})\nonumber\\
0 &= \left.\partial_w^m \partial_s^n \left(g_{\epsilon \epsilon \epsilon \epsilon}^{\<0,0,0,0\>^+}\right)\right|_{*}, && (\textrm{$m$ odd})\nonumber\\
0 &= \left.\partial_w^m \partial_s^n \left(g_{\sigma \sigma \epsilon \epsilon}^{\<0,0,0,0\>^+} + g_{\epsilon \sigma \sigma \epsilon}^{\<0,0,0,0\>^+}\right)\right|_{*}, && (\textrm{$m$ odd})\nonumber\\
0 &= \left.\partial_w^m \partial_s^n \left(g_{\sigma \sigma \epsilon \epsilon}^{\<0,0,0,0\>^+} - g_{\epsilon \sigma \sigma \epsilon}^{\<0,0,0,0\>^+}\right)\right|_{*}, && (\textrm{$m$ even})\nonumber\\
0 &= \left.\partial_w^m \partial_s^n \left(g_{\sigma \epsilon \sigma \epsilon}^{\<0,0,0,0\>^+}\right)\right|_{*}. && (\textrm{$m$ odd})
\end{align}

\section{Computing dependencies}
\label{app:dependencies}

In the build process described in section~\ref{sec:building}, we need to compute dependencies of each element of an SDP so as to minimize redundant work. For example, we must be able to determine the {\tt Block3d}'s that a given {\tt CompositeBlock} depends on, or the {\tt Block3d}'s and {\tt CompositeBlock}'s that a given  {\tt PartChunk} depends on. One possible approach is to write and maintain separate routines for constructing an element and for determining its dependencies. However, a downside of this approach is that the two routines must be manually kept in sync with each other, and this is error-prone. It is common to make changes to many parts of a computation when debugging, and it is furthermore common to spin off smaller example computations, such as an SDP for $T$-$\sigma$ correlators alone. We prefer not to have to worry about keeping track of dependencies during such changes.

\begin{lstlisting}[float, language=Haskell, caption=Example of using polymorphism in an {\tt Applicative} to simultaneously encode a computation and its dependencies. The comments show an example interactive session in {\tt ghci} using this code., label=lst:haskellExample]
{-# LANGUAGE ApplicativeDo #-}

type Key = String

myComputation :: Applicative f => (Key -> f Int) -> f Int
myComputation fetch = do
  x1 <- fetch "Hello"
  x2 <- fetch "Dolly"
  x3 <- fetch "Yes"
  pure (x1 + 2*x2 + 3*x3)

-- > import Data.Functor.Identity
-- > import Data.Functor.Const
-- > runIdentity $ myComputation (Identity . length)
-- 24
-- > getConst $ myComputation (\k -> Const [k])
-- ["Hello", "Dolly", "Yes"]
\end{lstlisting}

Instead we use a technique from \cite{DBLP:journals/corr/CapriottiK14} (nicely reviewed in \cite{10.1145/3236774}) that leverages Haskell's polymorphism to write a single function that simultaneously describes how to build an element and determine its dependencies. We first started using this method in \cite{Erramilli:2022kgp}, but to our knowledge it has not yet been described in the bootstrap literature.

A toy example of the technique is shown in listing~\ref{lst:haskellExample}. In this example, the routine {\tt myComputation} takes a function {\tt fetch} that returns an {\tt Int} for each {\tt Key} (here {\tt Key} is {\tt String}) in some {\tt Applicative} context {\tt f}. It calls {\tt fetch} on a few {\tt Key}s, and does a computation on the results. When we run {\tt myComputation} with {\tt f} instantiated to {\tt Identity} and {\tt fetch} the length function, it performs the computation on the lengths of the {\tt Key}s ``inside" {\tt myComputation}. However, at the same time, we can access the {\tt Key}s themselves by instantiating {\tt f} to {\tt Const [Key]}. In this case, no computation is performed, and only the {\tt Key}s are collected. Thus, because {\tt myComputation} is polymorphic in {\tt f}, it elegantly describes both a computation and the {\tt Key}s that the computation depends on.

 In a more realistic example, we could have a function {\tt fetchCompositeBlock} with type {\tt Applicative f => (BlockTableKey -> f Block3d) -> f CompositeBlock}. When the {\tt f} is instantiated to {\tt IO}, {\tt fetchCompositeBlock} can fetch {\tt Block3d}s from files and assemble them into appropriate linear combinations. By contrast, when {\tt f} is instantiated to {\tt Const [BlockTableKey]}, then {\tt fetchCompositeBlock} simply lists the {\tt BlockTableKey}s that are needed without doing any fetching or computation. This technique also composes nicely. We can use {\tt fetchCompositeBlock} as a subroutine in a more complicated function {\tt fetchSDP} with type {\tt Applicative f => (BlockTableKey -> f Block3d) -> f SDP}. This more complicated function is then capable of both building an {\tt SDP} and listing all the {\tt BlockTableKey}s that go into it, depending on how {\tt f} is instantiated.

In this way, we get the best of both worlds: we can pre-compute and memoize our dependencies, but the explicit construction routine is robust and nearly invisible to the end-user once the general {\tt fetch} routines are in place.

\section{Scheduler for building SDPs}
\label{app:scheduler}

In this appendix, we describe our ad-hoc scheduler for building SDPs. Recall from section~\ref{sec:building} that each SDP depends on several elements: some number of {\tt Block3d}'s, {\tt CompositeBlock}'s, {\tt PartChunk}'s, and {\tt Part}'s. From these elements, we build a tree of tasks, where an edge indicates a dependency: a child task must be completed before the parent task can be started. We label each task with a memory estimate, a runtime estimate function (which maps a number of cores $n$ to a time), and a minimum and maximum number of threads. (For example, our {\tt CompositeBlock, PartChunk, and Part} tasks are not parallelized, so they each have $\mathtt{maxThreads}=\mathtt{minThreads}=1$.) We assume that we have access to $K$ nodes, each with $C$ CPUs, and a fixed amount of memory $M$. Our algorithm has two phases:

\subsection{Phase 1: initial task distribution}
\label{sec:initialtaskdistribution}

To begin, we collect the tasks that have no dependencies that need to be built (which are usually {\tt Block3d} tasks), and sort them in reverse order of memory usage. We find a distribution of these initial tasks to the $K$ nodes by traversing the list in order. For a task $t$, we sort the nodes by how quickly they are expected to complete their current task assignments, and filter out nodes for which adding $t$ would decrease their ``score." We assign $t$ to the earliest-completing node whose score will be increased, and such that we will not exceed the memory of the node. When we encounter a $t$ such that there doesn't exist a node satisfying these requirements (for example, because none of the nodes have enough extra memory to accommodate $t$), the initial task distribution phase is over, and we proceed to the next phase.

The above algorithm depends on two definitions: a {\it completion time\/} and a {\it score\/} for a node with a given task assignment. We estimate {\it completion time\/} by finding a locally optimal allocation of CPUs to the assigned tasks that minimizes the maximum completion time of all assigned tasks (based on the runtime estimate function). We find this allocation in a simple way: we start with an even distribution of CPUs, and then steal CPUs from the fastest task and give them to the slowest task (while taking care not to violate the {\tt minThreads} and {\tt maxThreads} requirements), as long as overall performance increases. The {\it score\/} is a heuristic measure that we give to an allocation of CPUs to tasks, based on the idea that the completion times of the individual tasks should be similar.

After obtaining the initial task distribution, we start the corresponding tasks on their assigned nodes. 

\subsection{Phase 2: continually running tasks}

Once tasks are running, we switch to phase 2. We maintain a priority queue of tasks that are ready to run. In practice, we choose the priority to be a tuple $(\mathtt{depth},\,\mathtt{memory})$, where $\mathtt{depth}$ is the depth of the task in the tree (deeper tasks have higher priority), and $\mathtt{memory}$ is the task's memory estimate.

When a task finishes, we compute its reverse dependencies. If any of them are ready to be started (because all their child tasks have been completed), we add them to the priority queue. Each node keeps track of its number of free CPUs and amount of free memory and monitors the queue. If a node finds a task at the front of the queue that it can perform, given its free resources, it dequeue's that task. If it has enough resources, a node can dequeue multiple tasks at once, in which case it allocates free CPUs to them using the same algorithm as in subsection~\ref{sec:initialtaskdistribution}. It then begins running the newly dequeued tasks.

By running this algorithm, eventually all tasks will be completed.

\subsection{Collecting resource usage data}

The algorithm described above depends on a memory estimate and a runtime estimate for each task. When possible, we tried to build an analytical model predicting these resource requirements. This is painstaking because it requires intimate knowledge of how the task is performed, together with good measurements. These analytical models are needed when we perform a fresh build with new values of the parameters ($\ell_\mathrm{max}$, {\tt order}, etc. discussed in appendix~\ref{app:params}) that we have not tried before.

However, during every build, we record the actual resource usage of each task. Once we have this data, we can use it for memory and runtime estimates in subsequent builds. (This is possible because changing parameters like $\Delta_\sigma$ and $\Delta_\epsilon$ does not affect resource requirements.) Thus, after collecting resource usage data, we have accurate information about resource requirements, and we are no longer reliant on ad-hoc analytical models.

\section{Parameters}
\label{app:params}

We ran a number of tests changing different parameters in our computation to find the optimal parameters that yield accurate results without wasting too much CPU time. The parameters are $\ell_{max}$, giving the maximum spin included in our constraints, {\tt order}, giving the order of the radial expansion used in generating the conformal blocks, {\tt poles}, giving the effective number of poles used in the block interpolation described in~\ref{sec:interpolation}, and {\tt prec}, giving the internal binary precision used in our computations.

\begin{table}[t]
    \begin{tabular}{llllllll}
        $\ell_{max}$ & {\tt order} & {\tt poles} & {\tt prec} & primal objective & $\delta$ & $\delta^{(\ell_{max})}$ & $\delta^{(\ell_{max},\textrm{{\tt order}})}$\\
        \hline
        40 & 100 & 30 & 1088 & - (SDPB failed) & - & - & - \\
        \hline
        50 & 100 & 30 & 1088 & ${4.034436016}\times 10^{-8}$ & ${2.}\times 10^{-10}$ & 0 & 0\\
        \hline
        60 & 100 & 30 & 1088 & ${4.0256247689}\times 10^{-8}$ & ${1.1}\times 10^{-10}$ & 0 & 0\\
        \hline
        70 & 100 & 30 & 576 & - (SDPB failed) & - & - & - \\
        70 & 100 & 30 & 640 & ${4.0210005644}\times 10^{-8}$ & ${6.3}\times 10^{-11}$ & 0 & 0\\
        70 & 100 & 30 & 768 & ${4.0210005644}\times 10^{-8}$ & ${6.3}\times 10^{-11}$ & 0 & 0\\
        70 & 100 & 30 & 1088 & ${4.0210005644}\times 10^{-8}$ & ${6.3}\times 10^{-11}$ & 0 & 0\\
        \hline
        80 & 100 & 30 & 1088 & ${4.018319405}\times 10^{-8}$ & ${3.6}\times 10^{-11}$ & 0 & 0\\
        \hline
        88 & 95 & 80 & 1088 & - (SDPB failed) & - & - & -\\
        88 & 96 & 80 & 1088 & ${0.050659646253}$ & ${0.051}$ & ${0.051}$ & 0\\
        88 & 97 & 80 & 1088 & ${0.37298630331}$ & ${0.37}$ & ${0.37}$ & 0\\
        88 & 98 & 80 & 1088 & ${4.0176280315}\times 10^{-8}$ & ${2.9}\times 10^{-11}$ & ${6.}\times 10^{-12}$ & 0\\
        88 & 99 & 80 & 1088 & ${4.0253171472}\times 10^{-8}$ & ${1.1}\times 10^{-10}$ & ${8.3}\times 10^{-11}$ & 0\\
        88 & 100 & 20 & 1088 & ${3.9985768333}\times 10^{-8}$ & ${1.6}\times 10^{-10}$ & ${1.8}\times 10^{-10}$ & ${1.8}\times 10^{-10}$\\
        88 & 100 & 25 & 1088 & ${4.0169619513}\times 10^{-8}$ & ${2.2}\times 10^{-11}$ & ${6.5}\times 10^{-13}$ & ${5.2}\times 10^{-23}$\\
        88 & 100 & 30 & 1088 & ${4.0169619513}\times 10^{-8}$ & ${2.2}\times 10^{-11}$ & ${6.5}\times 10^{-13}$ & ${1.8}\times 10^{-24}$\\
        88 & 100 & 40 & 1088 & ${4.0169619513}\times 10^{-8}$ & ${2.2}\times 10^{-11}$ & ${6.5}\times 10^{-13}$ & ${3.1}\times 10^{-28}$\\
        88 & 100 & 60 & 1088 & ${4.0169619513}\times 10^{-8}$ & ${2.2}\times 10^{-11}$ & ${6.5}\times 10^{-13}$ & 0\\
        88 & 100 & 80 & 1088 & ${4.0169619513}\times 10^{-8}$ & ${2.2}\times 10^{-11}$ & ${6.5}\times 10^{-13}$ & 0\\
        88 & 101 & 80 & 1088 & ${4.0168942283}\times 10^{-8}$ & ${2.2}\times 10^{-11}$ & ${1.3}\times 10^{-12}$ & 0\\
        88 & 110 & 80 & 1088 & ${4.0170267048}\times 10^{-8}$ & ${2.3}\times 10^{-11}$ & ${1.2}\times 10^{-18}$ & 0\\
        88 & 120 & 80 & 1088 & ${4.0170267049}\times 10^{-8}$ & ${2.3}\times 10^{-11}$ & 0 & 0\\
        \hline
        100 & 100 & 30 & 1088 & ${4.0157281606}\times 10^{-8}$ & ${1.}\times 10^{-11}$ & 0 & 0\\
        \hline
        120 & 100 & 30 & 1088 & ${4.0147229801}\times 10^{-8}$ & 0 & 0 & 0\\
    \end{tabular}
    \caption{Accuracy tests for $\Lambda=35$ navigator.}
    \label{tab:accuracy-nmax18}
\end{table}

\begin{table}[t]
    \begin{tabular}{llllllll}
        $\ell_{max}$ & {\tt order} & {\tt poles} & {\tt prec} & primal objective & $\delta$ & $\delta^{(\ell_{max})}$ & $\delta^{(\ell_{max},\textrm{{\tt order}})}$\\
        \hline
        60 & 120 & 50 & 768 & ${2.565956605}\times 10^{-8}$ & ${1.6}\times 10^{-10}$ & 0 & 0\\
        \hline
        80 & 120 & 20 & 768 & ${2.5546288291}\times 10^{-8}$ & ${5.1}\times 10^{-11}$ & ${1.5}\times 10^{-11}$ & ${1.5}\times 10^{-11}$\\
        80 & 120 & 30 & 768 & ${2.5561529711}\times 10^{-8}$ & ${6.6}\times 10^{-11}$ & ${9.9}\times 10^{-17}$ & ${9.9}\times 10^{-17}$\\
        80 & 120 & 40 & 768 & ${2.5561529613}\times 10^{-8}$ & ${6.6}\times 10^{-11}$ & ${5.3}\times 10^{-20}$ & ${5.3}\times 10^{-20}$\\
        80 & 120 & 50 & 1024 & ${2.5561529613}\times 10^{-8}$ & ${6.6}\times 10^{-11}$ & ${1.9}\times 10^{-24}$ & ${1.9}\times 10^{-24}$\\
        80 & 120 & 60 & 768 & ${2.5561529613}\times 10^{-8}$ & ${6.6}\times 10^{-11}$ & 0 & 0\\
        \hline
        100 & 110 & 30 & 768 & - (SDPB failed) & - & - & - \\
        100 & 116 & 30 & 768 & ${0.8804118909}$ & ${0.88}$ & ${0.88}$ & 0\\
        100 & 118 & 30 & 768 & ${0.0033975568417}$ & ${0.0034}$ & ${0.0034}$ & 0\\
        100 & 120 & 50 & 768 & ${2.5518907348}\times 10^{-8}$ & ${2.3}\times 10^{-11}$ & ${2.}\times 10^{-13}$ & 0\\
        100 & 120 & 50 & 1024 & ${2.5518907348}\times 10^{-8}$ & ${2.3}\times 10^{-11}$ & ${2.}\times 10^{-13}$ & 0\\
        100 & 130 & 50 & 1024 & ${2.5519103657}\times 10^{-8}$ & ${2.4}\times 10^{-11}$ & ${8.}\times 10^{-18}$ & 0\\
        100 & 140 & 50 & 1024 & ${2.5519103665}\times 10^{-8}$ & ${2.4}\times 10^{-11}$ & ${9.6}\times 10^{-24}$ & 0\\
        100 & 150 & 50 & 1024 & ${2.5519103665}\times 10^{-8}$ & ${2.4}\times 10^{-11}$ & 0 & 0\\
        \hline
        120 & 120 & 50 & 1024 & ${2.5495483927}\times 10^{-8}$ & 0 & 0 & 0\\
    \end{tabular}
    \caption{Accuracy tests for $\Lambda=43$ navigator.}
\label{tab:accuracy-nmax22}
\end{table}

\begin{figure}[]
    \centering{}
    \includegraphics[width=0.7\paperwidth]{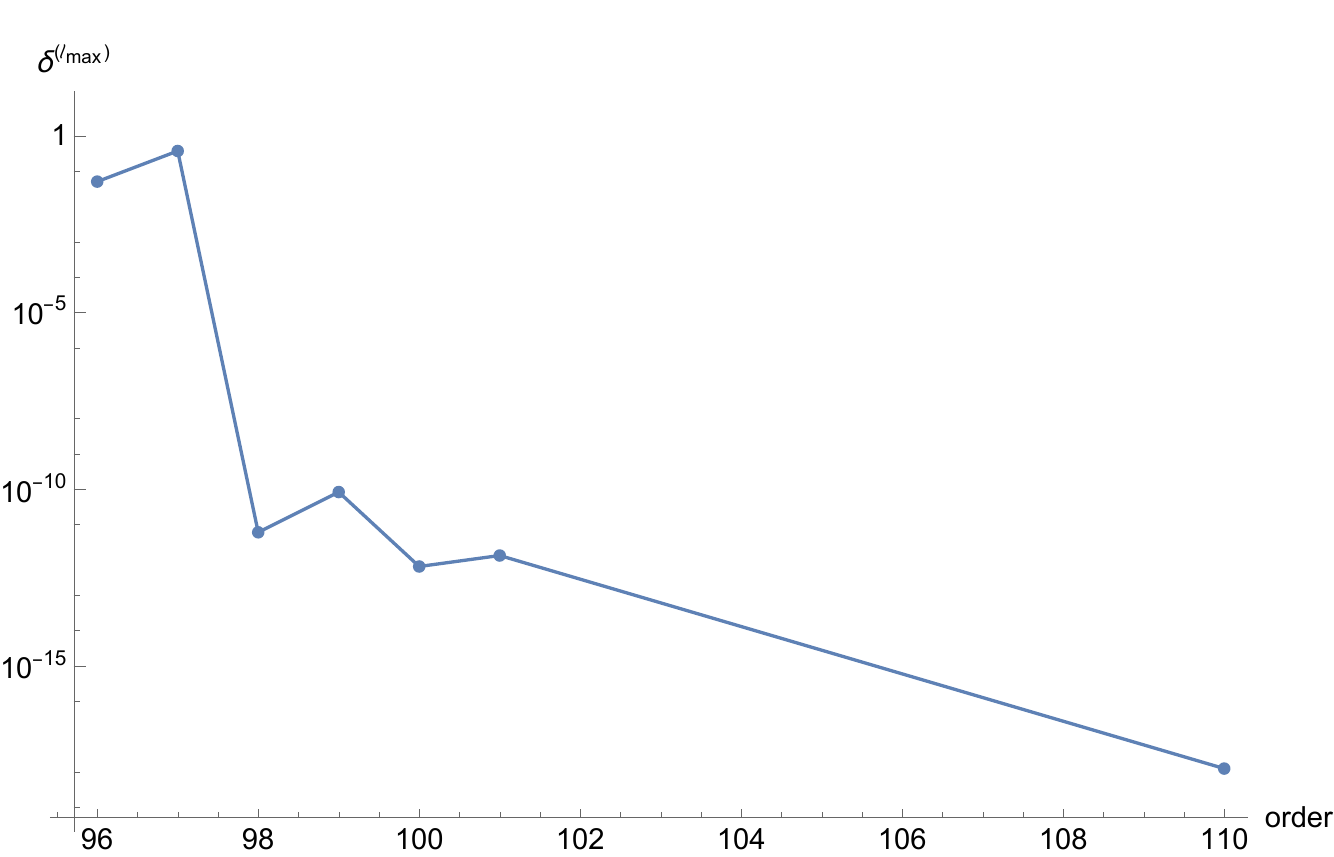}
    \caption{
        \label{fig:accuracy}
    Primal objective accuracy $\delta^{(\ell_{max})}$ at fixed $\ell_{max}=88$
    vs {\tt order}, $\Lambda=35$.
    }
\end{figure}

The results for $\Lambda=35$ and $\Lambda=43$ are summarized in tables~\ref{tab:accuracy-nmax18} and~\ref{tab:accuracy-nmax22}, respectively.
For a given set of spins $0 \ldots \ell_{max}$, {\tt order}, {\tt poles}, and {\tt prec}, we computed the primal objective defined as the value of the GFF navigator function.

Then, we computed the accuracy $\delta$ as the absolute difference of the primal objective and the most accurate primal objective (i.e.\ computed with the largest  $\ell_{max}$ and reasonable values of other parameters, see the last rows in tables~\ref{tab:accuracy-nmax18} and~\ref{tab:accuracy-nmax22}).
In a similar way, we computed the accuracy at a given $\ell_{max}$, denoted as $\delta^{(\ell_{max})}$, and
the accuracy at a given $\ell_{max}$ and {\tt order}, denoted as $\delta^{(\ell_{max},\textrm{{\tt order}})}$.

One can make the following observations from the accuracy tests:

\begin{itemize}
    \item Objective converges slowly with increasing $\ell_{max}$.
    This means that, after a certain point, increasing {\tt order} or number of poles for fixed $\ell_{max}$ does not improve accuracy (see the column $\delta$ in Tables~\ref{tab:accuracy-nmax18} and~\ref{tab:accuracy-nmax22}).
    \item There is a sharp cutoff at low {\tt order}s: for $\Lambda=35$, SDPB fails to converge at $\textrm{{\tt order}} =95$, converges to a completely wrong result for $\textrm{{\tt order}}=96$ and $\textrm{{\tt order}}=97$, and behaves well starting from $\textrm{{\tt order}}=98$ (see fig.~\ref{fig:accuracy} and the column $\delta^{(\ell_{max})}$ in table~\ref{tab:accuracy-nmax18}).
    $\Lambda=43$ exhibits similar behavior and converges to a correct result starting from $\textrm{{\tt order}}=120$ (see the column $\delta^{(\ell_{max})}$ in table~\ref{tab:accuracy-nmax22}).
    
    \item There is a cutoff for precision as well.
    For example, for $\Lambda=35$, $\text{{\tt prec}} = 576$ leads to SDPB crashing due to numerical errors.
    Any precision starting from 640 yields the same result.
    \item At fixed spins and {\tt order}, results converge rapidly with increasing number of poles.
    For example, for $\Lambda=35$, $\ell_{max}=88$, $\textrm{{\tt order}} = 100$, increasing the number of poles to any value beyond 25 will change the objective by less than $10^{-22}$ (see the column $\delta^{(\ell_{max},\textrm{{\tt order}})}$ in table~\ref{tab:accuracy-nmax18}).
\end{itemize}

Based on these test results, we chose the following parameters for our computations to ensure that the conformal block approximation and spin truncation induce a relative error which is negligible compared to other sources (e.g.~the error due to the Delaunay search):
\begin{itemize}
    \item $\Lambda=35$: $\ell_{max}=75$, $\textrm{{\tt order}}=106$, $\textrm{{\tt poles}}=35$, $\textrm{{\tt prec}}=832$.
    \item $\Lambda=43$: $\ell_{max}=90$, $\textrm{{\tt order}}=130$, $\textrm{{\tt poles}}=35$, $\textrm{{\tt prec}}=832$.
\end{itemize}

Let us emphasize in particular that it is important to systematically increase {\tt order} (the order of the radial expansion of conformal blocks used by {\tt blocks\_3d}) as we increase $\Lambda$. From our testing, we estimate that $\textrm{{\tt order}}\sim 5.5 \Lambda$ is roughly sufficient to ensure that errors remain small. This might seem surprising, given that the radial expansion for conformal blocks  \cite{Hogervorst:2013sma} is an expansion in $r=3-2\sqrt 2\approx 0.17$, which is numerically quite small. However, the coefficient of each power of $r$ is a polynomial in external dimensions, and the growth in degree of these polynomials plays an important role in determining the error from truncating the series. We encourage users of {\tt blocks\_3d} to keep this in mind and be sure to test the effects of {\tt order} on any given bootstrap problem.

\bibliographystyle{JHEP}
\bibliography{refs}
\end{document}